\title{\normalfont Extrapolating from a homologous series of oligomers to the infinite-mer: it's a long long way to infinity}
\author{
        {\em K. Razi Naqvi\/} \\
                Department of Physics,
        Norwegian University of Science and Technology\\
        NO-7491 Trondheim, Norway\\[1ex]
         }
\date{}

\documentclass[11pt,reqno]{article}
\usepackage{amsmath,amssymb,amsfonts} 
\usepackage{amssymb}  
\usepackage{makeidx}  
\usepackage{graphicx} 
\usepackage{epsfig}
\usepackage[sort&compress,numbers]{natbib}
\usepackage{pstricks}
\usepackage{soul}  
\usepackage[dvips]{curves}  
\usepackage{xcolor}
\usepackage{listings}
\usepackage{amsbsy}

\usepackage{multicol,caption}
\usepackage{float} 
\newenvironment{Figure}
  {\par\medskip\noindent\minipage{\linewidth}}
  {\endminipage\par\medskip}
\usepackage{pstricks, pst-text}                  
\usepackage{color}
\usepackage{fancyhdr}

\usepackage{mathpazo}
\usepackage{multicol,caption}
\usepackage{booktabs}
\makeatletter
\AtBeginDocument{
\renewcommand\NAT@open{\color{green}[}}
\usepackage{url}
\usepackage[bottom]{footmisc}
 2
 0
 1
 2
 1
 2
 2
 1
 1
 2
 0
 2
 1
 1
 2
 1
 2
 2
 1
 1
 2
 1

\definecolor{hellgrau}{gray}{.8}
\definecolor{dunkelblau}{rgb}{0, 0, .7}
\definecolor{roetlich}{rgb}{1, .7, .7}
\definecolor{dunkelmagenta}{rgb}{.3, 0, .3}
\definecolor{razi01}{rgb}{.3, .7, .3}
\definecolor{razi02}{rgb}{.7, .7, .3}
\definecolor{razi03}{rgb}{.1, .5, .5}
\definecolor{razi04}{rgb}{.5, .5, .5}
\definecolor{razi05}{rgb}{.6, .4, .6}
\definecolor{razi06}{rgb}{.1, .3, .6}

\long\def\symbolfootnote[#1]#2{\begingroup%
\def\thefootnote{\fnsymbol{footnote}}\footnote[#1]{#2}\endgroup} 

\graphicspath{%
    {converted_graphics/}
    {/}
    {Graphics/Chap13/}
    {Graphics/}
    {D:/}
}

\begin{document}

\maketitle

\begin{center}
\section*{\fontfamily{phv}\selectfont\normalsize Abstract}
\label{sec:Abstract}
\end{center}
\vspace{-1ex}

The usual strategy for deducing the $\pi\mbox{--}\pi^\ast$ electronic energy (or optical bandgap) in a molecule with an ``infinite" number of conjugated double bonds consists in fitting a function with some adjustable parameters to the relevant data for a set of homologous molecules with increasing number of repeat units ($N$), and assuming that, after its parameters have been optimized according to the least-squares criterion,  the function can be extended indefinitely, and its output will coincide with, or come close to, the correct limit.  Since more than ten homologues are seldom available, one might wonder whether extrapolation to the infinite-mer upon such slender basis is an instance of sound inductive reasoning or a mere leap of faith. The present article argues that the shape of the fitting function is an equally important criterion, and points out that the expressions proposed by Hirayama and by Meier and coworkers are flat functions of $1/N$, and that such functions are incongruent with the currently available evidence. Formulas derived by Davydov and by W. Kuhn are shown to be special cases of a new equation that outperforms both. \\[6ex]


\begin{multicols}{2}

\section{Introduction}

Few topics have exerted a greater influence, or commanded a more enduring interest, than the correlation between the length of a conjugated molecule and its electronic properties. The initial interest, ignited by the urge to puzzle out the relation between chemical constitution and colour \cite{LewisCalvin1939CR, Davydov1948JETP, Kuhn1948Helvetica,
 Kuhn1949JCP, Dewar1952JCS3, Hirayama1955JACS, Huzinaga1957PTP}, is sustained at present by the drive to fabricate molecular wires and other tailor-made components for nanodevices \cite{Tao2006NatureNanotech,Zade2010ACR}, and by the necessity to interpret the performance of conjugated polymers---systems that are ill-defined and polydisperse---in terms of the behaviour of well-characterized long oligomers.

A plethora of formulas have been proposed in an effort to relate  $\lambda_N$ (a suitably chosen wavelength in the most prominent absorption band of a conjugated oligomer) to $N$, the number of repeat units \cite{LewisCalvin1939CR, Davydov1948JETP, Kuhn1948Helvetica,Kuhn1949JCP, Dewar1952JCS3, Hirayama1955JACS, Huzinaga1957PTP, SdM1999, Meier1996Liebigs}, and three excellent reviews---two   \cite{Meier2005Ange,Gierschner2007AdvMat} dealing with general issues and the third   \cite{Torras2012JPCA} with numerical details---have appeared during the last decade. The authors of these contributions have focussed primarily on comparing experimental observations with the predictions of various formulas in their ``raw" forms, not allowing themselves to be distracted by (or benefit from) the fact that, more often than not, an algebraic formula can be manipulated so as to bring one or another of its features into prominence. As a result, expressions which are closely related have been treated as dissimilar rivals, like has been compared with unlike, the role of constraints, whether purposely enforced or inadvertently incorporated, has been overlooked, and equations which should have long been discarded, or never adopted in the first place, have continued to enjoy undeserved popularity.

This paper adopts a Procrustean approach by seeking to impose structural isomorphism on several formulas that have heretofore been regarded as bearing no particular relation to one another; it asks not how different the rival expressions are, but how closely they could be made to resemble one another, and how this imposed uniformity could be made to yield better results and provide, at the same time, a deeper understanding of the fitting formulas themselves.   

The rest of this paper is organized as follows. Section~2 begins with some comments (on the choice of the wavelength used for specifying the ``colour" of a conjugated molecule), which is followed by an explanation of the symbols that appear frequently in the following sections. Two three-parameter equations, both having the same form, are presented at the beginning of Section~3, and it is shown that four of the many special cases and subcases of these equations coincide with the formulas proposed by Hirayama \cite{Hirayama1955JACS}, by Meier and coauthors \cite{Meier1996Liebigs,  Meier1997ActaPol, Meier2002EJOC, Meier2005Ange}, by Davydov \cite {Davydov1948JETP} and by Werner Kuhn \cite{Kuhn1948Helvetica}. 
Section 4 will examine the equation of Hans Kuhn \cite{Kuhn1949JCP}, as well as its extensions \cite{SdM1999, Autschbach2007JCE}, and reasons will be given as to why the free electron model on which these equations are based should be regarded as an unviable option; two particular favourites, posited on the assumption that the transition energy bears a linear or quadratic relation with $Z=N^{-1}$, will also be mentioned in \S~\ref{subsection:EasyLazy}, but only very briefly, because Meier and coworkers (hereafter abbreviated as M\&Co) and other authors \cite{Gierschner2007AdvMat, Zade2006OrgLett} have shown that these fits cannot hold over a large range of $N$.

In what follows, I will shorten the {\em names\/} Nayler and Whting to N\&W, Hans Kuhn to HK, and Werner Kuhn to WK; the {\em equations\/} proposed by the last two and by Davydov will be called HaKu, WeKu, and DavY, respectively. A list of frequently used abbreviations has been provided in Appendix F.

\section{Symbols and notation}

The treatments published during the period 1948--52 were critically reviewed in 1955 by N\&W  \cite{Nayler1955JCS}, who stressed that it was not legitimate, as had been done by earlier authors, to compare theoretical predictions with data pertaining to a series consisting of simple examples among the shorter polyenes but highly alkylated examples of longer polyenes, a procedure which allowed ``a considerable amount of freedom in the selection of experimental values to suit a given theoretical expression".

In order to subject the formulas derived by WK, HK and Dewar to a rigorous test, N\&W synthesized a series of dimethyl polyenes with the general formula CH$_3$--[CH=CH]${_N}$--CH$_3$, with $N=\mbox{3--10 and 12}$, whose absorption spectra (in hexane and chloroform) revealed a well-resolved vibrational progression. They departed from HK and Dewar, who had picked $\lambda_{\rm max}^{[N]}$ (the wavelength corresponding to the peak absorbance) as the quantity of prime importance, and chose to work with $\lambda^{[00]}_{N}$ (the wavelength representing the first peak in the progression, justifying their choice in the following words: ``This [the 0-0 peak] has the practical advantage of being the sharpest and most easily distinguished, and the theoretical advantage that it must represent the zero-zero transition as regards the main vibrational processes in the newly excited molecule and therefore measures directly the energy difference in question". The use of $\lambda_{\rm max}^{[N]}$ is more practical when one is working with molecules whose spectra are devoid of vibrational structure, provided that one is certain that the longest wavelength band arises from a single electronic transition. 

Except when confusion is likely to arise or when the contrary is stated, I will use the simpler symbol $\lambda_N$ instead of $\lambda^{[00]}_{N}$ or $\lambda^{[\rm{max}]}_{N}$. Since it will be crucial to distinguish the measured value $\lambda_N$ from the corresponding predicted value, the symbol $\underline{\lambda}_N$ will be used for the latter quantity. At present, no more justification is needed for this fastidiousness than pointing out that, if different symbols are not used for the two wavelengths, it will not be possible to define the difference between the two as $\delta_N\equiv\lambda_N - \underline{\lambda}_N$, and the squared sum of errors as 
\begin{equation}\label{eq:DefDelta}
\Delta \equiv \sum_{\scriptsize\mbox{all $N$}}\delta_N^2 .
\end{equation}
The author begs leave to assure the reader that there is more to his insistence on a crisp notation than mere pedantry, that its neglect amounts to ignoring the difference between a constrained and an unconstrained fit, to comparing (unwittingly or surreptitiously) oranges with apples. 

In the formulas given below, $X$ will invariably stand for the parameter whose optimized value is to be interpreted as the best estimate for ${\lambda}_N$ provided by a given fitting function, and $Y\equiv X^{-1}$ will represent the corresponding estimate for the limiting energy; wavelengths will {\em always\/} be measured in nm, but the units for energy $E=C_0/\lambda$ will be either $\mbox{cm}^{-1}$ ($C_0=10^7$) or eV ($C_0=10^4/8.0655$). The predicted value for the energy will be denoted by $\underline{E}_N$; the measured value, by $E_N$; since the symbol $\Delta$ will be retained when energy is used instead of wavelength, the unit for $\Delta$ will not be explicitly indicated.

\section{New equations for old}

We will be interested, for the most part, in fitting functions that can be expressed as
\begin{align}
\underline{\lambda}_N &= X\left[1-K \Xi(k,N)\right ]^{q}, \label{eq:MasterX}\\
\noalign{\noindent\mbox or as }
\underline{\lambda}_N &= X\left[1-A \Theta(a,N)\right ]^{q}, \label{eq:MasterT}
\end{align}
where $K$ and $k$ (and $A$ and $a$) are adjustable parameters, and $q$ is restricted to one of two values ($1 \mbox{ or } {\textstyle\frac{1}{2}}$).

The functions $\Xi$ (exponential kernel) and $\Theta$ (trigonometric kernel) are defined below:
\begin{align}
\Xi  & =\exp(-kN),\label{eq:ExpKer}\\
\Theta & = \frac{1-\cos\theta_N}{1-a\cos\theta_N}.\label{eq:TrigKer}
\end{align}

A concrete versions of eq~\ref{eq:MasterX} will be called X3pU or X3pH according as $q=1$ or $q=\textstyle\frac{1}{2}$; likewise, the two concrete versions of eq~\ref{eq:MasterT} will be called T3pU and T3pH. The first character identifies the kernel (X for exponential, T for trigonometric), the last character identifies the value of $q$ (unity or half), and the middle two specify the number of free parameters.

If one sets $\Xi=\exp(-kN)$ and $q=\textstyle\frac{1}{2}$ in eq~\ref{eq:MasterX}, it becomes equivalent to the relation
\begin{equation}\label{eq:Hirayama01}
\underline{\lambda}_N = \left (\alpha-\beta \gamma^{N} \right )^{1/2}, \quad (\gamma<1),
\end{equation}
proposed by Hirayama \cite{Hirayama1955JACS}, who used a different notation.

Let us pause to observe that eqs~\ref{eq:MasterX} and \ref{eq:MasterT} can each be reduced to a two-parameter equation by the imposition of a constraint; in labeling the reduced version of a three-parameter equation, the numerical character will be changed from 3 to 2. The process of reduction will be illustrated by considering a special case ($q=1$) of eq~\ref{eq:MasterX}, namely the equation
\begin{equation}\label{eq:X3pU1}
\underline{\lambda}_N = X\left[1-K e^{-kN}\right ].
\end{equation}
Setting $N=s$ in eq~\ref{eq:X3pU1}, where $s$ is a suitably chosen constant, and solving for $K$, one gets
\begin{equation}\label{eq:Kequ}
K=e^{ks}\left (1-\frac{\underline{\lambda}_s}{X} \right).
\end{equation}
If we impose the demand $\underline{\lambda}_s = \lambda_s$, introduce a new symbol $K_s$ through the relation
\begin{equation}
e^{ks}\left (1-\frac{{\lambda}_s}{X} \right)=e^{ks}K_s,
\end{equation}
and replace $K$ by $K_s e^{ks}$ on the right-hand side of eq~\ref{eq:X3pU1} (X3pU), we arrive at X2pU, the two-parameter variant of X3pU,
\begin{equation}
\underline{\lambda}_N  = X\bigl [1-K_s{e}^{-k(N-s)}\bigr ], \label{eq:X2pU}\\
\end{equation}
which can also be expressed as
\begin{equation}\label{eq:X2pFb}
\underline{\lambda}_N =\underline{\lambda}_\infty - (\underline{\lambda}_\infty-\lambda_s){e}^{-k(N-s)}.
\end{equation}

Equation~\ref{eq:X2pFb} subsumes the equation proposed by Meier and coauthors (M\&Co), who confined themselves to the special case $s=1$; they did not use different symbols for the measured and calculated wavelengths, and overlooked, or failed to point out, that their equation was a three-parameter equation in disguise. M\&Co demonstrated that their equation provided excellent fits to many data sets \cite{Meier1996Liebigs,Meier1997ActaPol,Meier2002EJOC,Meier2005Ange}.

Now, M\&Co, who proposed eq~\ref{eq:X2pFb} (with $s=1$), {\em also\/} recommended, as an {\em independent\/} recipe for computing the energy, the relation
\begin{equation}\label{eq:MeierE}
\underline{E}_N = Y - (Y - E_1){\rm e}^{-\epsilon\hspace{0.1ex}(N-1)},
\end{equation}
which will henceforth be called Meier's equation for energy (MfE). Evidently, MfE and the equation (cf. eqs~ \ref{eq:X2pU} and \ref{eq:X2pFb})
\begin{subequations}\label{eq:X2pUE}
\begin{align}
\underline{E}_N & =  \left [ \underline{\lambda}_\infty - (\underline{\lambda}_\infty-\lambda_1){\rm e}^{-k(N-1)}\right ]^{-1} \label{eq:X2pUEa}\\
&=Y \left[ 1 - (1-Y/E_1){\rm e}^{-k(N-1)}\right ]^{-1}, \label{eq:X2pUEb}
\end{align}
\end{subequations}
cannot both provide equally good fits to the same data set. When the two equations, MfE and eq~\ref{eq:X2pUEa}, which are an algebraically incongruent pair, are applied to the same spectral data, one must find that the relations $\underline{E}_\infty=C_0 /\underline{\lambda}_\infty$ and $\epsilon=k$ are not obeyed; M\&Co did perform such a comparison  \cite{Meier1996Liebigs,Meier1997ActaPol}, but did not comment on the discrepancies. So far as the experience of the present author goes, eq~\ref{eq:X2pUE} provides a closer fit than MfE, which means that the latter should be discarded.

Of the two concrete versions of eq~\ref{eq:MasterX}, X3pH is not new, but each equation with the trigonometric kernel (T3pU/H) is new, and so are their two-parameter versions.

Two fitting formulas from the past (both using two parameters) will now be recalled and stated, in keeping with the general theme of this article, in a single relation,
\begin{equation}\label{eq:2pTriG}
\underline{\lambda}_N= W(1-A\cos\theta_N)^{-q}
\end{equation}

If we put $q=\textstyle\frac{1}{2}$ in eq~\ref{eq:2pTriG}, we recover W. Kuhn's formula \cite{Kuhn1948Helvetica}, whereas the other option ($q=1$) gives Davydov's formula \cite{Davydov1948JETP} for the polyphenyls, which was derived by using a simplified quantum mechanical description analogous to that used by W. Kuhn. It is easily verified that eq~\ref{eq:MasterT} reduces to eq~\ref{eq:2pTriG} when one sets $a=A$, and uses the relation $X(1-A)^q=W$, which follows from eq~\ref{eq:2pTriG}.

The reader's attention should also be directed to yet another earlier relation based on a trigonometric function, even though it could not be coerced into adopting the form of eq~\ref{eq:MasterT}. Using a quantum chemical approach, Huzinaga and Hasino \cite{Huzinaga1957PTP} arrived at a two-parameter formula (to be called H\&H) that can be stated as follows:
\begin{equation}\label{eq:HandHa}
\underline{\lambda}_N  = 2X \left [1+\sqrt{1+P\sin^2(\pi/N_\pi)}\, \right ]^{-1},
\end{equation}
where $N_\pi$ is the number of $\pi$ electrons.

It is shown in Appendix B that DavY, H\&H and WeKu can each be reduced, through the imposition of a constraint, to a one-parameter version. Appendix D compares the performance of WeKu and H\&H.

\subsection{FEMO model} \label{subsection:FEMO}

Using the free electron molecular orbital (FEMO) approach, HK \cite{Kuhn1949JCP} deduced a one-parameter relation (HaKu),
\begin{equation}\label{eq:HK1p}
\underline{E}_N = Y \left (1-\frac{1}{N_\pi}\right ) + \frac{h^2}{8mL^2} (N_\pi+1),
\end{equation}
where $h$ is Planck's constant, $m$ is the electronic mass, $L$ is the length of the conjugated chain (the distance between the points where where the potential can be assumed to be infinite, which HK took to be one bond distance to either side of the terminal C-atoms of the conjugated chain system), and $Y$ has the same meaning as in earlier expressions.

HaKu has been extended by turning it into a three-parameter relation of the form shown below \cite{SdM1999}:
\begin{equation}\label{eq:HK3p}
\underline{E}_N = Y \left [1-\frac{1}{N_\pi}+ \frac{U}{(N_\pi+l)^2} (N_\pi+1)\right ].
\end{equation}

We note for later purposes that, upon expressing the right-hand side of eq~\ref{eq:HK3p} as a series in $1/N_\pi$ and substituting $N_\pi=2N^{=} = pN$, one gets, after setting
\begin{align}
c_1 &=(U-1)/p, \quad c_2= U(1-2l)/4p,\\
\noalign{\vspace{1ex}}
\noalign{\noindent\mbox the result shown below: }
\noalign{\vspace{1ex}}
\underline{E}_N &= Y( 1 + c_1N^{-1} + c_2N^{-2} + \cdots ).\label{eq:HK3pSer}
\end{align}
It is noteworthy that the parameter $l$ appears only in the second-order term, which means that, even if its introduction improves the fit at small values of $N$, it plays no role in determining the limiting value $\underline{E}_\infty$.

Further analysis of Eqs.~\ref{eq:HK1p}--\ref{eq:HK3pSer} has been consigned to Appendices D and E.

\subsection{``EeZy" and ``LaZy" plots} \label{subsection:EasyLazy}

A common device to display the approach to saturation is to plot $\underline{\lambda}_N$ (or $\underline{E}_N$) against $N$. Despite their simplicity and directness, such plots cannot easily communicate differences in the large-$N$ behaviour of the predicted curves; however, this disadvantage can be easily overcome by using $Z\equiv N^{-1}$ as the abscissa. Since an $E$ {\em versus\/} $Z$ plot is frequently used as an easy means of estimating the limiting energy $\underline{E}_\infty$ (see below), it seems appropriate to designate such a representation as an ``EeZy plot". However, there being no compelling theoretical argument against using the wavelength for the same purpose, it is convenient to coin the term ``LaZy plot" for a graph showing the variation of $\lambda$ with $Z$; it need hardly be added that the new words are to be pronounced like ``easy" and ``lazy".

One reason for the popularity of EeZy plots is the widespread opinion that the large-$N$ behaviour of $E_N$ may be satisfactorily described by the equation
\begin{equation}\label{eq:Linear}
\underline{E}_N =Y + \frac{\mbox{constant}}{N},
\end{equation}
which will be called PN2p (polynomial with two parameters) or the linear relation. Various modifications of the linear relation, including that shown below,
\begin{equation}\label{eq:Poly02}
\underline{E}_N = Y + A_1 N^{-1}+ A_2 N^{-2},
\end{equation}
have also been used for fitting spectral data. Equation \ref{eq:Poly02} will be called PN3p.

It is not hard to fathom why the practice of representing $\underline{E}_N$ as a polynomial (of first or second degree) in $N^{-1}$
was adopted in the days when nonlinear regression analysis was beyond the reach of most workers, but its continuing allure in the face of incontrovertible evidence defies understanding. 

\subsection{Asymptotic behaviour}

As previous works have paid scant attention to the large-$N$ behaviour of fitting functions (other than those based on polynomials in $Z$), it is necessary to fill this gap here.

To understand the large-$N$ behaviour of curves generated by functions with exponential kernels, it is helpful to recall the definition of a {\em flat function\/}.\footnote{A function $f(x)$ is said to be flat at $x_0$ if it is infinitely differentiable at $x_0$ and all its derivatives at this point vanish. This implies that a formal Taylor expansion of $f(x)$ at $x=x_0$ does not exist. Geometrically, this means that the line $y=x_0$ has a contact of infinite order with the curve of the function at the point $x_0$ \cite[p.~462]{Courant1965IntroCalc}. The Arrhenius relation $k=A\exp(-E/RT)$ is an example of a flat function \cite{Glaister1991MathGaz}.} The four functions which use an exponential kernel, being flat functions, appear (on the scale of ordinary plots, such as those shown here) to become parallel to the horizontal axis well before meeting the vertical axis. In contrast, the large-$N$ behaviour of functions with a trigonometric kernel can be expressed as (see Appendix C) shown below:
\begin{subequations}\label{eq:Large-N}
\begin{align}
\underline{\lambda}_N & = X - \alpha Z^{2},\\
\noalign{\medskip}
\underline{E}_N & = Y + \beta Z^{2}.
\end{align}
\end{subequations}
where $\alpha$ and $\beta$ are constants ($>0$) satisfying the relation $\alpha/X =\beta/Y$. 

The contrasting behaviours of functions with the exponential and trigonometric kernels bear a close analogy to two well-known models of the specific heat of solids.\footnote{Einstein's expression for the specific heat of a solid \cite[p.~214]{Slater1939IntroChemPhys} is a flat function at $T=0$, whereas Debye's theory predicts a $T^3$ dependence in the limit $T\to 0$ \cite[p.~236]{Slater1939IntroChemPhys}.}

\begin{Figure} 
  \centering
\includegraphics[width=\textwidth]{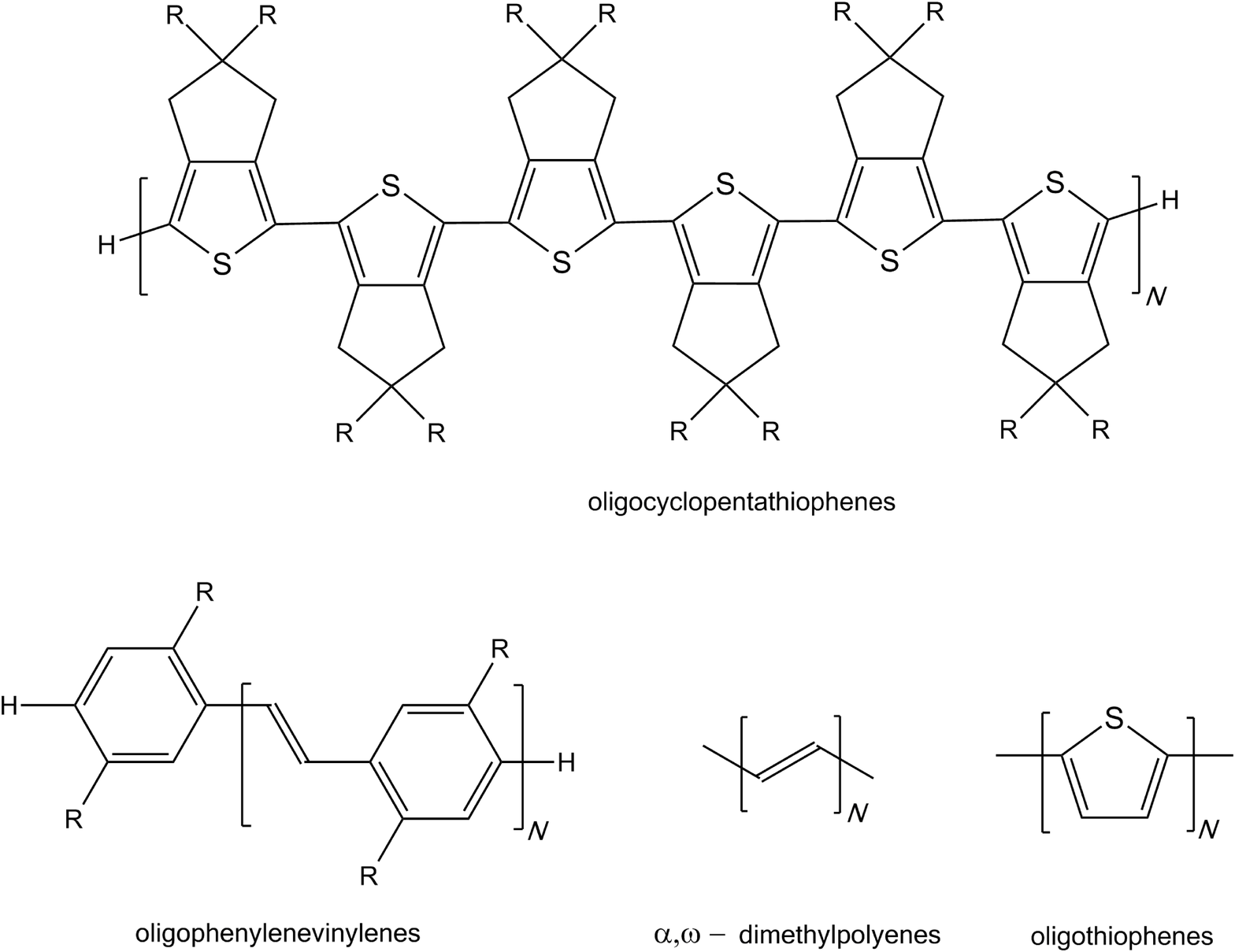}
  \captionof{figure}{Molecular structures of the oligomers mentioned in the text}
  \label{Figure:FigureOne}
\end{Figure}


To find out which, if either, of the large-$N$ behaviours mentioned above conforms with experimental observations, one needs a data set that is sufficiently dense and precise. Fortunately, a trove of spectroscopic data satisfying these stringent requirements has been made available by Izumi and coauthors \cite{Izumi2003JACS}, who synthesized  a series of oligothiophenes ($N=2, 4, 6, 12, 18, 24, 36, 48, 72, 96$; see Fig.~\ref{Figure:FigureOne}). They have listed both $\lambda_N^{[{\rm max}]}$ for their (unstructured) absorption spectra and $\lambda_N^{[00]}$ for their (structured) fluorescence emission spectra; since the precision of their absorption measurements is stated to be 0.2 nm for absorption and 0.4 nm for emission spectra, I have chosen their absorption data for further inspection. Plots resulting from some two-parameter fits are displayed in Figure~\ref{Figure:IzumiEeZy}; in order to be able to focus attention on the large-$N$ behaviour, the data for the shorter molecules ($N=2,4,6$) 
have not been included in these plots, but Table~\ref{tab:IzumiTab} displays the energies of all oligomers.  {\em On the whole\/} (that is to say, when the entire set is considered), excellent fits ensue from X2pU/H, T2pU/H and WeKu, but T2pU and T2pH stand out in particular (and are indistinguishable on the scale of the plot). The reason for the inferior performance of flat functions is easy to understand if one divides the oligomers into two subsets, one for which $N\leq 24$  and another for which $N\geq 36$, and notes that $\delta_N$ is positive (negative) for one set and negative (positive) for the other. Curves generated by flat functions (or exponential kernels), where saturation sets in abruptly (between $N=24$ and $N=36$), minimize the sum of squared residuals ($\Delta=\sum_N\delta_N^2$) by changing the sign of $\delta_N\equiv \underline{\lambda}_N-{\lambda}_N$ in this region, which produces systematic deviations on either side of the divide.
\begin{Figure} 
  \centering
\includegraphics[width=+.5\textwidth]{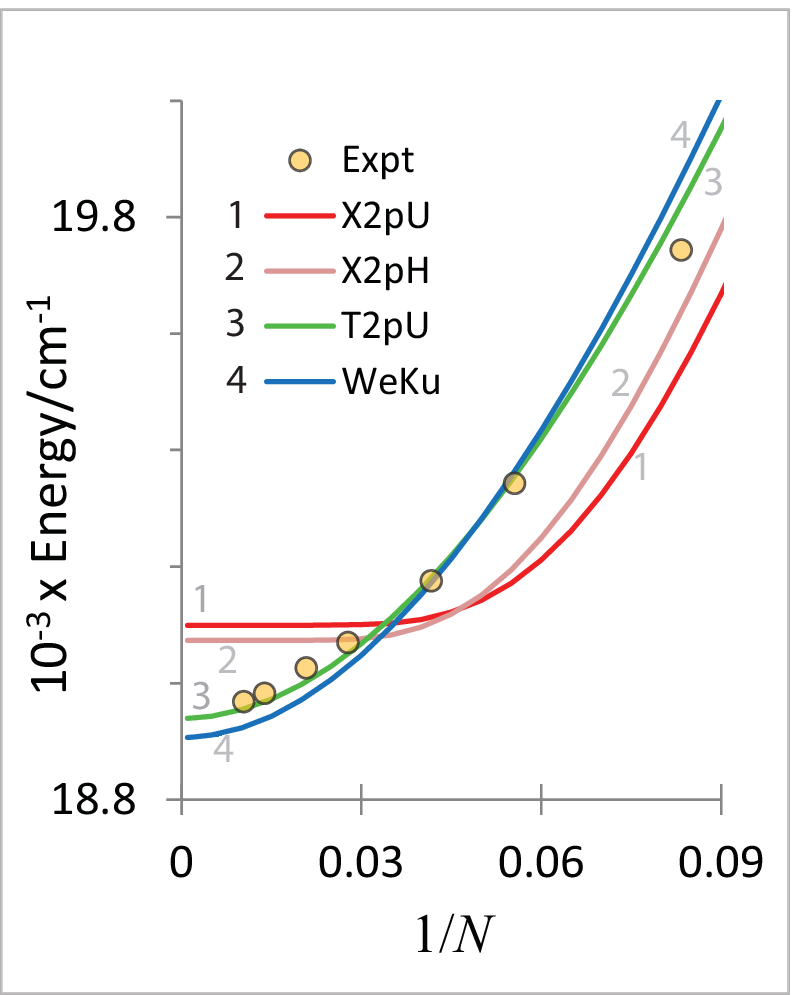}  
  \captionof{figure}{Experimental values ($\lambda_N$) and predicted wavelengths ($\underline{\lambda_N}$) of maximal absorbance for cycolpentathiophene oligomers synthesized by Izumi and coworkers \cite{Izumi2003JACS}.}
  \label{Figure:IzumiEeZy}
\end{Figure}

\begin{Figure} 
  \centering
\includegraphics[width=+.5\textwidth]{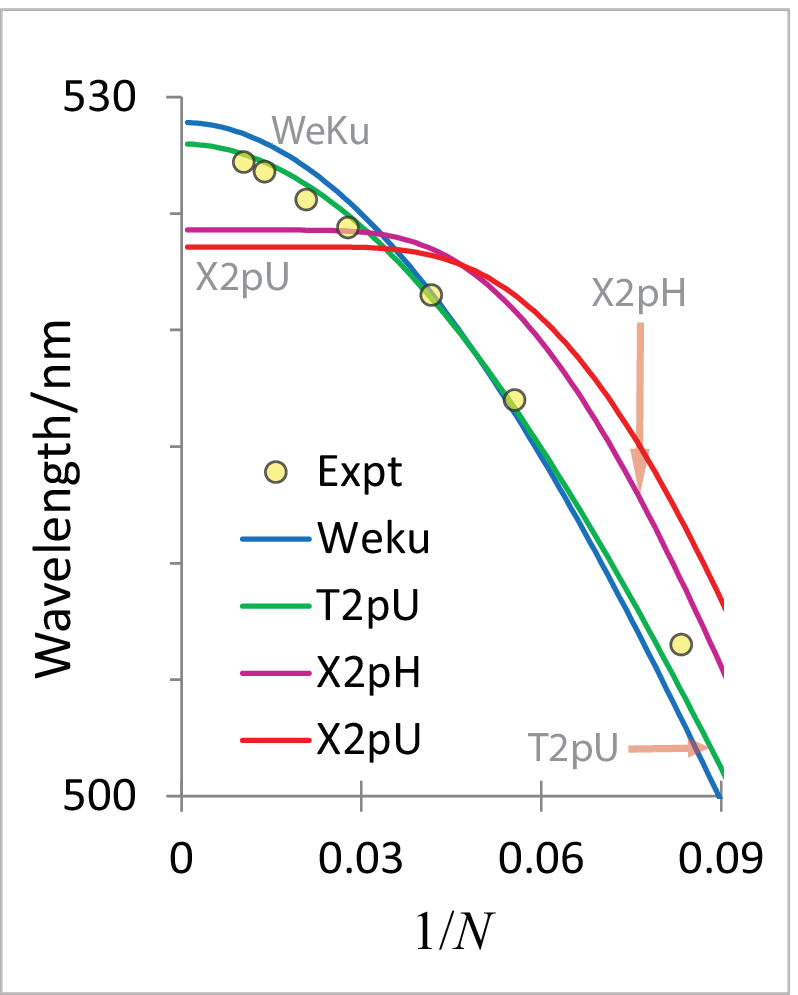} 
  \captionof{figure}{Observed and predicted values of the energy ($E_N^{[\rm max]}$ and $\underline{E}_N^{[\rm max]}$, respectively) for cycolpentathiophene oligomers synthesized by Izumi and coworkers \cite{Izumi2003JACS}.}
  \label{Figure:IzumiLaZy}
\end{Figure}

Similar behaviour, but less pronounced than that discussed above in relation to Figures \ref{Figure:IzumiEeZy} and \ref{Figure:IzumiLaZy}, can be discerned if one plots the data published by M\&Co \cite{Meier1997ActaPol, Meier2002EJOC,Meier2005Ange}. The structure of their oligophenylenevinylenes is shown in Figure~\ref{Figure:FigureOne}, and $N=1,2,\ldots 8, 11, 15$; following the nomenclature of M\&Co, these molecules will be labeled {\bf 1a}--{\bf 1j}.

\begin{Figure} 
  \centering
\includegraphics[width=0.55\textwidth]{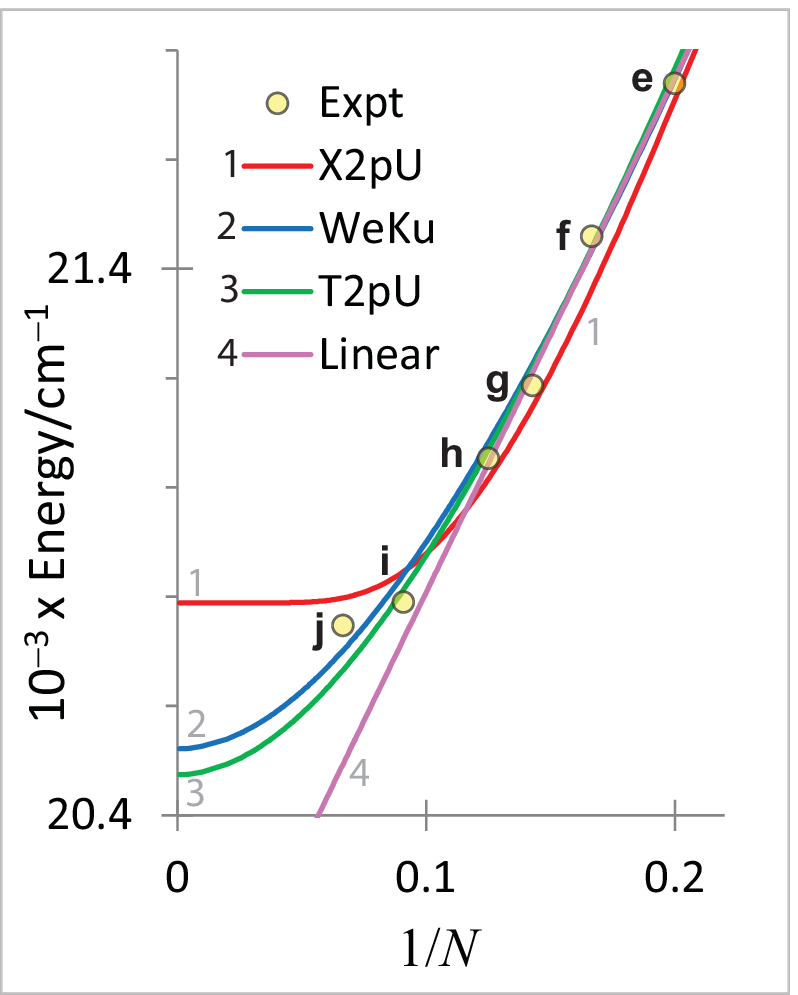}
  \captionof{figure}{``EeZy" plots showing $E_{N}^{\rm max}$ (symbols) for $N=5\mbox{--}8, 11, 15$ (compounds {\bf 1e}--{\bf 1j}) and the values predicted by four formulas (curves);  for other details, see the text.}
  \label{Figure:MeierFig1}
\bigskip
\end{Figure}

Figure~\ref{Figure:MeierFig1} is an extended rendering of Figure~1 in Refs.~\cite{Meier2005Ange} and \cite{ Meier2002EJOC}, where the abscissa is taken as $1/N^\ast$ and energy is measured in eV; $N^\ast=N+1$ is the number of benzene units. M\&Co also synthesized polyphenylenevinylenes with different end groups (the so-called red polymers {\bf 1p}, {\bf 1p'}, and {\bf 1p''}, with $\lambda_{\rm max} =482, 477$, and 485 nm, respectively).  They attributed the smaller value ($\lambda_{\rm max}=477$ nm) for {\bf 1p'} to the 
presence of shorter segments, the longer value ($\lambda_{\rm max}=485$ nm) for {\bf 1p''} to the induction of a push-pull character by the imino endgroup, and plotted the point for {\bf 1p} in their figure as a confirmation of their estimate $\underline{\lambda}_\infty = 481$ nm. Since M\&Co tested only two fitting functions, and found one, the linear fit, to be patently inappropriate, their conclusion (that, of the three possible candidates for a surrogate infinite-mer, only {\bf 1p} fitted the bill) is not altogether inevitable. There remains the possibility that both {\bf 1p} and {\bf 1p'} were adulterated with shorter chains, and that {\bf 1p''}  might well be a good substitute for a really long ($N\approx 30$) oligophenylenevinylene. If we discard the ambiguous results pertaining to the polymers, the oligomer data are consistently closer to WeKU and T2pU than to X2pU. At least two more oligomers (with $N>15$) and spectral measurements with four significant digits are needed for a satisfactory resolution of the issue.

{\small
\begin{Figure}
\begin{minipage}{0.9\linewidth}
\renewcommand{\arraystretch}{1.2}
\renewcommand\footnoterule{}
\begin{center}
\captionof{table}{Experimental and predicted values of $\lambda_N^{[{\rm max}]}$ (in nm) of oligocyclopentathiophenes (structure shown in Figure~\ref{Figure:FigureOne}) of Izumi {\em et al.\/}  \cite{Izumi2003JACS}} 
\medskip
\label{tab:IzumiTab}
\begin{footnotesize}
\begin{tabular}{rlcccc}
\toprule\\[-3ex]
&\multicolumn{1}{c} {\normalsize$\lambda_N$} &\multicolumn{3}{c} {{\normalsize$\underline{\lambda}_N$}} \\  \cmidrule(r){2-2} \cmidrule(r){3-5}
$N$	&Expt	&T2pU	&T2pH	&WeKU\\
\hline\hline
2	&313.0	&313.0	&313.0	&316.4\\
4	&408.5	&409.5	&409.9	&407.4\\
6	&457.5	&457.2	&457.0	&454.4\\
12	&506.5	&504.6	&504.3	&503.4\\
18	&517.0	&516.7	&516.6	&516.5\\
24	&521.5	&521.4	&521.4	&521.7\\
36	&524.4	&525.0	&525.1	&525.6\\
48	&525.6	&526.3	&526.4	&527.1\\
72	&526.8	&527.3	&527.4	&528.1\\
96	&527.7	&527.6	&527.8	&528.5\\
\\[-2.75ex]
\hline\\[-2.75ex]
&{\footnotesize $\underline{\lambda}_\infty$}    	&528.1	&528.2	&529.0	 \\
&{\footnotesize $\Delta$}\hspace{1ex}  			&6.0	&8.9	&38.6\\
&{\footnotesize $R^2$}\hspace{1ex}  			&0.9999	&0.9998	&0.9992\\
\\[-2ex]
\toprule
\end{tabular}
\end{footnotesize}
\end{center}
\end{minipage}
\end{Figure}
}

For a second series that is {\em just\/} long enough ($2\leq N \leq 50$) to discriminate between flat functions and those which are not flat, I turn to some quantum chemical (QC) calculations, reported to be within about 0.2 eV of the experimental value, of band gaps of oligothiophenes \cite{Zade2006OrgLett}. For the present purpose, it is not so much the accuracy of the computed energies as their dependence on $N$ that is of cardinal concern, and I will use the symbol $E_N$ for the QC energies reported by Zade and Bendikov \cite{Zade2006OrgLett}, because these data will serve as the benchmark for testing various fitting equations; the values of $E_N$ were reported, in accordance with standard practice in the field, to only two decimal places.

\begin{Figure} 
  \centering
\includegraphics[width=0.5\textwidth]{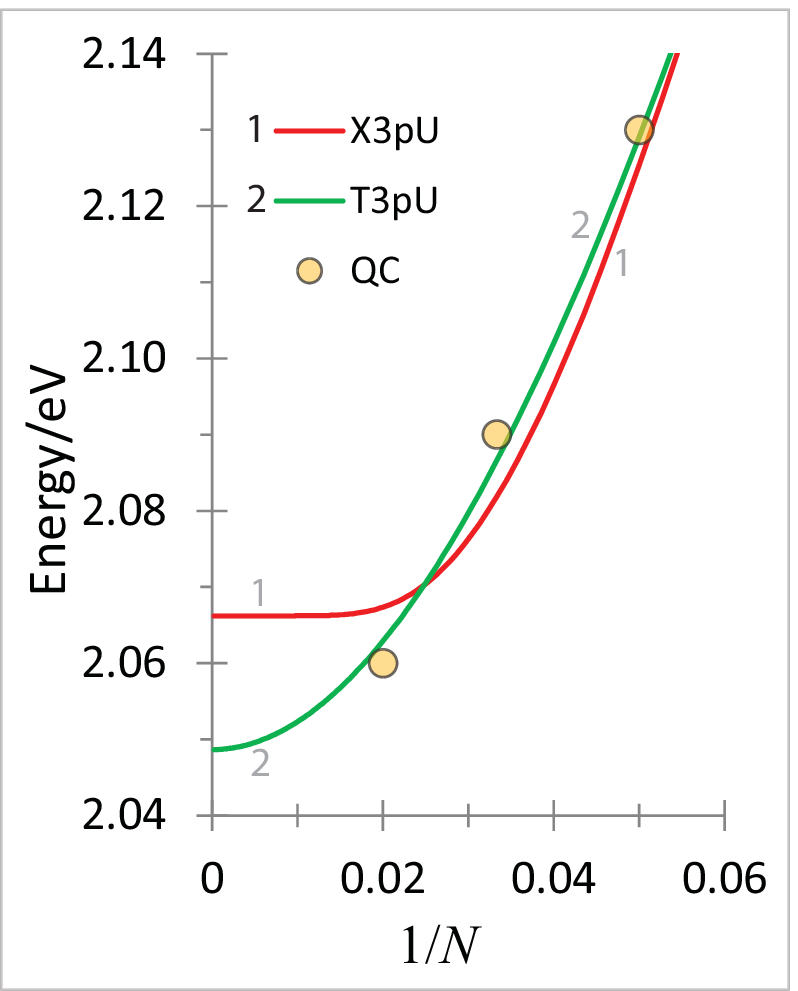}
  \captionof{figure}{Quantum chemical (symbols) and predicted values (curves) of the band gap in oligothiophenes; the QC data are from Ref.~\cite{Zade2006OrgLett} .}
  \label{Figure:ZadeQC}
\end{Figure}

Zade and Bendikov \cite{Zade2006OrgLett} made the interesting observation that, for the longer oligomers ($N\geq 10$), their values of $E_N$ could be very well represented ($R^2=0.9996$) by a second degree polynomial in $Z$ ($E_N=a_0+a_1Z+a_2Z^2$), and they deduced, by identifying $a_0$ with $\underline{E}_\infty$, the result $\underline{E}_\infty=2.03$ eV. When $\underline{E}_N$ was calculated (by the present author) by fitting the four three-parameter equations to the QC results ($E_{N}$) for the longer thiophenes ($10 \leq N \leq 50$), the values listed in the last three columns of Table~\ref{tab:Thio3pB} were obtained; since T3pH and T3pU gave identical results, reproducing $E_{N}$, and the outputs of X3pU/H never differ by more than 0.001 eV from $E_{N}$, the last character (H or U) in the label will be dropped temporarily. That $\underline{E}_\infty(\mbox{X3p})$ turns out to be 2.07, i.e. {\em larger\/} than ${E}_{50}$, constitutes, together with the plot in Fig.~\ref{Figure:ZadeQC}, a cogent argument for claiming that X3p fails to furnish a faithful approximation to $E_{N}$ at large values of $N$ ($N\to\infty$). 

\end{multicols}
\hrule
\begin{table}[ht]    
\begin{minipage}[b]{0.45\linewidth}\centering
\caption{Quantum chemical (QC) values of HOMO-LUMO gaps (in eV) of oligothiophenes  and four three-parameter fits}
\medskip
\label{tab:Thio3pB}
\begin{frame}{}
\begin{footnotesize}
\begin{tabular}{rlcccc}
\toprule\\[-2.5ex]
&\multicolumn{1}{c} {\small$E_N$} &\multicolumn{3}{c} {{\small$\underline{E}_N$}} \\  \cmidrule(r){2-2} \cmidrule(r){3-5}
$N$	&QC  	&T3pU/H &X3pU  &X3pH\\
\hline\hline\\[-2ex]
10	&2.31	&2.308	&2.304	&2.305\\
11	&2.27	&2.272	&2.272  &2.272\\
12	&2.24	&2.243	&2.244  &2.244\\
13	&2.22	&2.219	&2.221  &2.221\\
14	&2.20	&2.199	&2.201  &2.201\\
15	&2.18	&2.182	&2.183  &2.183\\
20	&2.13	&2.129	&2.125  &2.126\\
30	&2.09	&2.087	&2.082  &2.082\\
50	&2.06	&2.063	&2.067  &2.067\\
\\[-2ex]
\hline\\[-2ex]
{$\underline{E}_\infty$}    &2.03\footnote{Obtained through polynomial extrapolation}	&2.05\;\;	&2.07\;\; &2.07\;\;	 \\
{$\Delta$}\hspace{1ex}  	&		&4.5E$-$5 	&2.1E$-$4 &1.9E$-$4
\\[-1ex]
\\[-1ex]
\toprule
\end{tabular}
\end{footnotesize}
\end{frame}
\end{minipage}
\hspace{1.2cm}
\begin{minipage}[b]{0.45\linewidth}
\centering
\caption{Extrapolation of the fits in \mbox{Table~\ref{tab:Thio3pB}} (optimized for $10\leq N \leq 50)$) to shorter homologues}
\medskip
\label{tab:Thio3pA}
\begin{frame}{}
\begin{footnotesize}
\begin{tabular}{rlcccc}    
\toprule\\[-2.5ex]
&\multicolumn{1}{c} {\small$E_N$} &\multicolumn{4}{c} {{\small$\underline{E}_N$}} \\  \cmidrule(r){2-2} \cmidrule(r){3-6}
	$N$&QC  &T3pU	&T3pH	&X3pU  &X3pH\\
\hline\hline \\[-2ex]
2	&4.23	&3.19	&3.24	&2.93	&3.04 \\
3	&3.45	&2.99	&3.01	&2.79	&2.85 \\
4	&3.03	&2.82	&2.83	&2.67	&2.71 \\
5	&2.78	&2.68	&2.68	&2.58	&2.60 \\
6	&2.62	&2.56	&2.56	&2.5	&2.51 \\
7	&2.50	&2.47	&2.48	&2.44	&2.45 \\
8	&2.42	&2.41	&2.41	&2.39	&2.39 \\
9	&2.35	&2.35	&2.35	&2.34	&2.34
 \\
\toprule
\end{tabular}
\end{footnotesize}
\end{frame}
\vspace{2.0cm}
\end{minipage}
\end{table}
\vspace{-4mm}
\hrule
\begin{multicols}{2}

How rapidly a good fit deteriorates as one moves away from either end of the segment over which the function was fitted is a question of considerable importance. Each fit of Table~\ref{tab:Thio3pB} was therefore extended to shorter oligomers, and the results, displayed in Table~\ref{tab:Thio3pA}, show that even though T3p (and X3p) predict values of $\underline{E}_N$ that accord exactly (and almost exactly) with $E_N$ ($N\geq 10$), they do not enjoy the same success when the fits are extrapolated to shorter oligomers. Clearly, the reliability of an extrapolated result depends not on how far, in terms of $N$, the target of extrapolation is from the upper or lower end of the segment used for optimization, but on how far it is on the energy scale. The difference $E_2-E_{10}>2\mbox{ eV}$, whereas $E_{10}-E_\infty$ cannot be larger than 0.3 eV (assuming that ${E}_\infty \geq 2.03$ eV).

The situation of the greatest practical interest is that when one wishes to find $\underline{\lambda}_N$ after one has acquired spectral data for short and moderately long oligomers. Zade and Bendikov \cite{Zade2006OrgLett}, who used the symbol $n$ instead of $N$, found that the energies of the shorter thiophenes (up to $N\approx 12$) followed closely a linear dependence on $Z$, but they immediately added a caution that can hardly be overstressed: ``However, the linear prediction is not particularly accurate, yielding band gaps of 1.81 eV for polythiophene". They went on to draw a conclusion that requires a closer examination: ``Using short oligomers (up to $n=10$, as used in most previous studies) is insufficient for prediction of the band gaps of conducting polymers".

\end{multicols}
\hrule 
\begin{table}[h!]
\begin{minipage}[b]{0.45\linewidth}\centering
\caption{Extrapolation of three-parameter fits optimized for the shorter thiophenes ($2\leq N \leq 9)$) to longer homologues}
\medskip
\label{tab:Thio3pAll}
\begin{frame}{}
\begin{scriptsize}
\begin{tabular}{rlcccc}
\toprule\\[-2.5ex]
&\multicolumn{1}{c} {\small$E_N$} &\multicolumn{4}{c} {{\small$\underline{E}_N$}} \\  \cmidrule(r){2-2} \cmidrule(r){3-6}
	$N$&QC  &T3pU	&T3pH	&X3pU  &X3pH\\[0.5ex]
\hline\hline\\[-2ex]
2	&4.23	&4.23	&4.23	&4.23	&4.23 \\
3	&3.45	&3.46	&3.45	&3.45	&3.44 \\
4	&3.03	&3.03	&3.03	&3.03	&3.03 \\
5	&2.78	&2.77	&2.78	&2.78	&2.79 \\
6	&2.62	&2.61	&2.61	&2.61	&2.62 \\
7	&2.50	&2.50	&2.50	&2.50	&2.50 \\
8	&2.42	&2.42	&2.42	&2.42	&2.42 \\
9	&2.35	&2.36	&2.36	&2.36	&2.35 \\
\hline
10	&2.31	&2.32	&2.31	&2.32	&2.30\\
11	&2.27	&2.29	&2.28	&2.29	&2.26\\
12	&2.24	&2.26	&2.25	&2.26	&2.22\\
13	&2.22	&2.24	&2.23	&2.25	&2.20\\
14	&2.20	&2.23	&2.21	&2.24	&2.18\\
15	&2.18	&2.21	&2.20	&2.23	&2.16\\
20	&2.13	&2.17	&2.15	&2.20	&2.11\\
30	&2.09	&2.14	&2.12	&2.20	&2.08\\
50	&2.06	&2.12	&2.10	&2.20	&2.08\\
\hline\\[-2ex]
{$\underline{E}_\infty$}    &2.03	&2.11&2.09&2.20&2.08 
\\
\toprule
\vspace{4mm}
\end{tabular}
\end{scriptsize}
\end{frame}
\end{minipage}
\hspace{1.25cm}
\begin{minipage}[b]{0.45\linewidth}
\centering
\caption{QC results of energy of oligothiophenes  \cite{Zade2006OrgLett} and three two-parameter fits to the complete data set ($2\leq N \leq 50$)}
\medskip
\label{tab:ZadeWeKu}
\begin{frame}{}
\begin{scriptsize}
\begin{tabular}{rccccc}
\toprule\\[-2.5ex]
&\multicolumn{1}{c} {\small$E_N$} &\multicolumn{3}{c} {{\small$\underline{E}_N$}} \\  \cmidrule(r){2-2} \cmidrule(r){3-5}
$N$	&QC	&T2pU	&T2pH	&WeKU\\[0.5ex]
\hline\hline\\[-2ex]
2	&4.23	&4.23	&4.23	&4.20\\
3	&3.45	&3.48	&3.46	&3.47\\	
4	&3.03	&3.04	&3.04	&3.06\\	
5	&2.78	&2.78	&2.78	&2.79\\	
6	&2.62	&2.60	&2.61	&2.62\\	
7	&2.50	&2.49	&2.49	&2.50\\	
8	&2.42	&2.41	&2.41	&2.42\\	
9	&2.35	&2.34	&2.35	&2.35\\	
10	&2.31	&2.30	&2.30	&2.30\\	
11	&2.27	&2.26	&2.27	&2.27\\	
12	&2.24	&2.24	&2.24	&2.24\\	
13	&2.22	&2.22	&2.22	&2.21\\	
14	&2.20	&2.20	&2.20	&2.19\\	
15	&2.18	&2.18	&2.18	&2.18\\	
20	&2.13	&2.14	&2.13	&2.13\\	
30	&2.09	&2.10	&2.10	&2.09\\	
50	&2.06	&2.09	&2.08	&2.07\\	
\hline\\[-2ex]
&{$\underline{E}_\infty$}	&2.07		&2.07 		&2.06	 \\
&{$\Delta$}\hspace{1ex}    	&2.7E$-$3 	&1.1E$-$3 	&2.8E$-$3\\
&{$R^2$}\hspace{1ex}    	&0.9999 	&0.9998 	&0.9995
\\
\toprule
\end{tabular}
\end{scriptsize}
\end{frame}
\end{minipage}
\end{table}
\vspace{-3mm}
\hrule
\begin{multicols}{2}

When a long data string is available, one can always use an initial segment of the string for fitting purposes, and the remainder for checking how well a given fitting equation accounts for the (longer) oligomers not included in the fitting; the results of this enquiry, which took into reckoning oligothiophenes with $N\leq 9$, appear in Table~\ref{tab:Thio3pAll}. One sees that all four functions perform almost equally well, the difference $|\delta_N|\equiv |\lambda_N-\underline{\lambda}_N|$ never exceeding 0.01 eV, not only for the range of $N$ over which the fit was carried out but also for $N=10$. It will be assumed hereafter that $E_\infty=2.04\pm 0.01\mbox {eV}$. Saturation  sets in too early (and at a value about 0.15 eV larger than $E_\infty$) in X3pU, a little later in X3pH, and at a value rather close to $E_\infty$. The outputs of two-parameter fits are not shown here, but the interested reader can verify that the result of each two-parameter fit, though influenced a little by the choice of $s$, does not differ greatly from the output of its three-parameter parent. It goes without saying that if the final result of a two-parameter fit had turned out to depend sensitively on the choice of $s$, such sensitivity would have undermined the very idea of reducing the number of parameters from three to two.

Before considering a second example of using the data for an initial segment as a basis for predicting $\underline{\lambda}_\infty$,  it is important the stress that the precision of the data at hand sets a limit on the possibility to distinguish between rival formulas. The truth of this remark may be seen by examining Table~\ref{tab:ZadeWeKu}, which shows the results obtained by applying T2pU, T2pH and WeKu to the full data set for the band gaps of oligothiophenes; X2pU and X2pH were not included in this comparison because functions with the exponential kernel always give a poorer performance when they are applied to long data sets (where saturation is clearly in evidence). As it is not possible to discriminate, on the basis of the QC data alone, between WeKu and T2pU/H, it would be interesting to repeat the calculations if more precise QC data become available in the future; pending the availability of such calculations, the only recourse is to squeeze some more information from data that have already been analyzed.

{\small
\begin{Figure}
\begin{minipage}{0.9\linewidth}
\renewcommand{\arraystretch}{1.2}
\renewcommand\footnoterule{}
\begin{center}
\captionof{table}{Values of $\underline{\lambda}_N$ (in nm) for ``virtual" oligocyclopentathiophenes ($N=2\mbox{--}13$)} 
\medskip
\label{tab:IzumiVirt}
\begin{footnotesize}
\begin{tabular}{c||cccccccccc}
\toprule\\[-3.5ex]
$N$		&2	&3	&4	&5	&6	& 7\\
$\lambda_N$	&313.0	&366.6	&408.5	&438.8	&457.5	&473.6\\
\hline
$N$		&8	&9	&10	&11	&12 	&13\\
$\lambda_N$	&484.7	&492.7	&498.6	&503.0	&506.5	&509.2\\
\toprule
\end{tabular}
\end{footnotesize}
\end{center}
\end{minipage}
\end{Figure}
}
One option available at present is to return to Table~\ref{tab:IzumiTab}, which has a preponderance of long oligomers, and to use interpolation for producing data for shorter, ``virtual" oligomers. One set that satisfied the requirement that $\lambda_N$ must coincide with the original data (for $N=2,4,6,12,18$) was produced by fitting a second-degree polynomial (in $N$) to the spectral data for $N=2,4,6$ and a second-degree polynomial in $1/N$ to data for $N=6,12,18$, is shown in Table~\ref{tab:IzumiVirt}. A different procedure for interpolation is not expected to yield data differing by more than the uncertainty in the original data.

Table~\ref{tab:IzumiTab2} shows the results found by analyzing $\lambda_N^{[{\rm max}]}$ of Izumi and coworkers, and two segments ($N=2\mbox{--}10$ and $N=2\mbox{--}12$) from the data for ``virtual" oligocyclopentathiophenes. The data in the last four columns will be discussed first. If one declares the best fit to be that with the smallest value of $\Delta$, one must conclude that X3pU and X2pU outperform all the other fits. However, one should recall that, according to Table~\ref{tab:IzumiTab}, $\lambda_\infty$ must be very close to 528 nm, and, since the very purpose of making a fit is to deduce the value of $\underline{\lambda}_\infty$ from a data set that does not contain sufficiently long oligomers, one must conclude that X3pU and X2pU are, in fact particularly poor fits, because  they predict a value for $\underline{\lambda}_\infty$ that is close to 514 nm. Indeed, just how inadequate the flat functions are becomes obvious, as has already been stated, when one examines the fits to the full data set of ``real" oligomers (columns 2 and 3 in Table~\ref{tab:IzumiTab2}), and notices that all four fits with exponential kernels have much larger values of $\Delta$.\\

{\small
\begin{Figure}
\begin{minipage}{0.9\linewidth}
\renewcommand{\arraystretch}{1.2}
\renewcommand\footnoterule{}
\begin{center}
\captionof{table}{Predicted values of $\underline{\lambda}_\infty$ (in nm) and $\Delta$ for oligocyclopentathiophenes (see Figure~\ref{Figure:FigureOne}) of Izumi {\em et al.\/}  \cite{Izumi2003JACS}} 
\medskip
\label{tab:IzumiTab2}
\begin{footnotesize}
\begin{tabular}{ccrcrcrc}
\toprule\\[-3ex]
&\multicolumn{2}{c} {all} &\multicolumn{2}{c} {{$N=$2--10}}&\multicolumn{2}{c} {$N=$2--12} \\  \cmidrule(r){2-3} \cmidrule(r){4-5}\cmidrule(r){6-7}
$N$	&$\underline{\lambda}_\infty$	&$\Delta$&$\underline{\lambda}_\infty$	&$\Delta$&$\underline{\lambda}_\infty$	&$\Delta$\\
\hline\hline
T3pU		&528.1	&5.9		&534.2	&6.0		&532.7	&8.9\\
T3pH		&528.2	&8.8		&535.8	&8.4		&533.6	&13.0\\
X3pU		&523.7	&91.6		&513.8	&5.5		&514.1	&5.9\\
X3pH		&524.4	&48.4		&522.0	&17.1		&519.6	&19.9\\\hline
T2pU		&528.1	&6.0		&534.4	&6.0		&532.7	&9.1\\
T2pH		&528.2	&8.9		&536.2	&9.0		&533.9	&14.1\\
X2pU		&523.7	&93.7		&514.1	&5.8		&514.1	&6.1\\
X2pH		&524.4	&48.8		&522.9	&19.2		&520.2	&23.0\\\hline
WeKu		&529.0	&38.6		&538.2	&13.0		&536.3	&25.2\\
\toprule
\vspace*{-3ex}
\end{tabular}
\end{footnotesize}
\end{center}
\end{minipage}
\end{Figure}
}

Table~\ref{tab:IzumiTab2} shows that, in dealing with the data for ``virtual" oligomers, T2pU and T2pH provide a closer fit than WeKu (as judged by the values of $\Delta$), and make a better prediction for $\underline{\lambda}_\infty$.

\subsection{Concluding remarks}

It has been argued above that minimization of $\Delta$ (the sum of squared errors) should not be seen as the be all and end all when one is fitting a function to the spectral data ($\lambda_N$ or $E_N$) of a homologous series of conjugated molecules, especially if one's aim is to extrapolate the data to the corresponding infinite-mer, and that the shape of the fitting function should also be taken into account. Many new fitting functions have also been presented and shown to provide better fits than can be obtained by the formulas which have been used so far.\\[-1ex]

\noindent
{\bf Acknowledgment}\\[-2ex] 

I thank Hans-Richard Sliwka for keeping alive my interest in the topic through discussions stretching over more than a decade. One sometimes hears the proverb ``It is better to travel hopefully than to arrive". Under the prevalent reward system, few scientists can hope to prosper if they follow such a precept, but he sees scientific activity as a reward in itself. The amicable completion of this article has proved (at least to each of us) that it is possible to travel together hopefully {\em and\/} to arrive alone.\\
\end{multicols}

\hrule

\begin{center}
{\Large\bf Appendices A--G}
\end{center}
\vspace*{0.5cm}

\leftskip = 2cm
\rightskip= 2cm

\begin{minipage}{0.75\linewidth}
\hspace*{4cm}
\begin{description}
  \item[Appendix A:]Concrete forms of eqs~\ref{eq:MasterX} and \ref{eq:MasterT} \hfill p.~\pageref{page:AppA} 
  \item[Appendix B:]One-parameter versions of DavY, H\&H and WeKu \hfill p.~\pageref{page:AppB} 
  \item[Appendix C:]Limiting forms ($Z\to 0$) \hfill p.~\pageref{page:AppC}
  \item[Appendix D:]${\alpha}$-oligothiophenes \hfill pp.~\pageref{page:AppDstart}--\pageref{page:AppDfinish}
  \item[Appendix E:]A closer look at FEMO \hfill pp.~\pageref{page:AppEstart}--\pageref{page:AppEfinish} 
  \item[Appendix F:]{\em p\/}-Polyphenyls \hfill pp.~\pageref{page:AppFstart}--\pageref{page:AppFfinish}
  \item[Appendix G:]List of frequently used abbreviations and\\ symbols related to the length of an oligomer \hfill p.~\pageref{page:AppG}    
\end{description}

\leftskip = 0cm
\rightskip= 0cm

\end{minipage}
\newpage

\leftskip = 0cm
\rightskip= 0cm

\noindent
{\bf Appendix A: Concrete forms of eqs~\ref{eq:MasterX} and \ref{eq:MasterT}}\\\label{page:AppA}

\noindent Adjustable parameters are shown in red colour.\\[4ex]

\noindent
{\em Exponential kernel} (eq~\ref{eq:MasterX})
\begin{align}
q=1: \underline{\lambda}_N &= {\color{red}X} \left [1-{\color{red}K}{\rm e}^{-{\color{red}k}(N-s)}\right ]\tag{X3pU}\\
\noalign{\vspace{0.6ex}}
\underline{\lambda}_N &= {\color{red}X} \left[1-K_s{\rm e}^{-{\color{red}k}(N-s)}\right ], &K_s&=1-{\lambda}_s/{\color{red}X} \tag{X2pU}\\
\noalign{\vspace{1.4ex}}
q=\textstyle\frac{1}{2}:\underline{\lambda}_N &= {\color{red}X} \left [1-{\color{red}K}{\rm e}^{-{\color{red}k}(N-s)}\right ]^{1/2}\tag{X3pH}\\
\noalign{\vspace{0.6ex}}
\underline{\lambda}_N &= {\color{red}X} \left [1-K_s{\rm e}^{-{\color{red}k}(N-s)}\right ]^{1/2}, &K_s&=1-({\lambda}_s/{\color{red}X})^2 \tag{X2pH}
\end{align}

\selectfont
\noindent
Comments on special cases:
\begin{enumerate}
  \item X3pU is a new equation.
  \item X2pU (with $s=1$) was proposed earlier by M\&Co \cite{Meier1996Liebigs,Meier1997ActaPol,Meier2002EJOC,Meier2005Ange}.
  \item X3pH (with $s=0$) should be credited to Hirayama \cite{Hirayama1955JACS}.
  \item X2pH is a new equation.
\end{enumerate}

\vspace*{4ex}
\noindent
{\em Trigonometric kernel} (eq~\ref{eq:MasterT})

\begin{equation}
\left [ \mbox{Notation:}\quad \theta_N=\frac{\pi}{N+1}, \quad f(a,\theta_N)=\frac{1-\cos\theta_N}{1-a\cos\theta_N}\right ]
\end{equation}
\begin{align}
q=1: \underline{\lambda}_N &= {\color{red}X}\bigl [1-{\color{red}A}f({\color{red}a},\theta_N)\bigr ] \tag{T3pU}\\
\noalign{\vspace{0.6ex}}
\underline{\lambda}_N &=  {\color{red}X}\bigl  [1-A_sf({\color{red}a},\theta_N)\bigr ], &A_s&=\frac{1-({\lambda}_s/{\color{red}X})}{f({\color{red}a},\theta_s)} \tag{T2pU}\\
\noalign{\vspace{1.4ex}}
q=\textstyle\frac{1}{2}:\underline{\lambda}_N &= {\color{red}X}\bigl [1-{\color{red}A}f({\color{red}a},\theta_N)\bigr ]^{1/2} \tag{T3pH}\\
\noalign{\vspace{0.6ex}}
\underline{\lambda}_N &= {\color{red}X}\bigl  [1-A_sf({\color{red}a},\theta_N)\bigr ]^{1/2}, &A_s&=\frac{1-({\lambda}_s/{\color{red}X})^2}{f({\color{red}a},\theta_s)} \tag{T2pH}
\end{align}

\noindent
Comment: T3pU, T2pU, T3pH and T2pH are {\em all\/} new. \\

The four equations with the exponential kernel (X3pU, X3pH, X2pU, X2pH) will be called \mbox{\bf X-equations}; likewise, the four equations with the trigonometric kernel (T3pU, T3pH, T2pU, T2pH) will be called {\bf T-equations}.

\newpage

\noindent
{\bf Appendix B: One-parameter versions of DavY, H\&H and WeKu}\\\label{page:AppB}

\noindent{\em DavY and Weku}

The two-parameter formula (previously numbered as eq~\ref{eq:2pTriG})
\begin{subequations}
\begin{align}
\underline{E}_N       & = \hspace{0.6ex}F\left [1-A\cos\theta_N\right ]^{q} \hspace{1.2ex}=\underline{E}_1\left [1-A\cos\theta_N\right ]^{q},\\
\noalign{\vspace{1ex}}
\noalign{\noindent\mbox or its twin }
\noalign{\vspace{1ex}}
\underline{\lambda}_N & = W\left [1-A\cos\theta_N\right ]^{-q} =\underline{\lambda}_1\left [1-A\cos\theta_N\right ]^{-q},
\end{align}
\end{subequations}
in which $\underline{E}_N=1/\underline{\lambda}_N$ and $F=1/W$, is rather special because of the occurrence of $\underline{E}_1$ (or $\underline{\lambda}_1$) on the right-hand side. The equations proposed by Davydov and WK follow upon setting $q=1$ and $q=\textstyle\frac{1}{2}$, respectively.

\begin{enumerate}
  \item If the series happens to start with $N=1$, one can replace  $\underline{\lambda}_1$ with $\lambda_1$ (or $\underline{E}_1$ with $E_1$) and adjust the value of the remaining parameter, namely $A$, to minimize $\Delta$.
  \item When $N=1$ is not a part of the series, $W$ (or its inverse $F$) must be treated as an adjustable parameter, but $A$ can be replaced by $A_s$, defined by the relation
\begin{equation}
A_s=\frac{1}{\cos\theta_s}\left [1-\left (\frac{W}{\underline{\lambda}_s}\right )^{1/q} \right ]
   =\frac{1}{\cos\theta_s}\left [1-\left (\frac{\underline{E}_s}{F}\right )^{1/q} \right ].
\end{equation}
\end{enumerate}

\noindent{\em H\&H}

For this equation, only the second option is available. In terms of the angle
\begin{equation}
\phi_N=\frac{\pi}{N_\pi},
\end{equation}
the parameter $P$ in eq~\ref{eq:HandHa} can be replaced by $P_s$, which is defined below:
\begin{equation}
P_s=\frac{1}{\sin^2\phi_s}\left [\left (\frac{X}{2\underline{\lambda}_s}-1\right )^2 -1\right ]
=\frac{1}{\sin^2\phi_s}\left [\left (\frac{2\underline{E}_s}{Y}-1\right )^2 -1\right ].
\end{equation}

Appendix D shows the results obtained by using the reduced versions of WeKu and H\&H, which have been named WK1 and HH1, respectively.

\newpage

\noindent
{\bf Appendix C: Limiting forms ($Z\to 0$)} \\\label{page:AppC}

So far the symbols $A$ and $a$ have been used for all formulas using trigonometric kernels, but it would be safer to abandon this practice now. Accordingly, I will spell out each general formula again when its small-$Z$ expansion is given again. We begin by considering two cases of eq~\ref{eq:MasterT}.

\begin{align}
\mbox{Case 1 $(q=1)$}:\hspace{8ex} \underline{\lambda}_N &=\hspace{3ex} X\left [1-\,\frac {A(1-\cos\theta_N)}{1-a\cos\theta_N} \right ]   \hspace{8ex}\\
\noalign{\vspace{0.6ex}}
&\xrightarrow[Z\to 0]{}X\left [1-\,{\frac {A\pi^{2}}{2(1-a)}}{Z}^{2}+ {\cal O}(Z^{3}) \right ]\\
\noalign{\vspace{1.4ex}}
\mbox{Case 2 $(q=\textstyle\frac{1}{2})$}:\hspace{8ex} \underline{\lambda}_N &=\hspace{3ex} X\left [1-\,\frac {B(1-\cos\theta_N)}{1-b\cos\theta_N} \right ]^{1/2}\hspace{8ex}\\
\noalign{\vspace{0.6ex}}
&\xrightarrow[Z\to 0]{}X\left [1-\,{\frac {B\pi^{2}}{4(1-b)}}{Z}^{2} + {\cal O}(Z^{3})\right ]
\end{align}

\vspace{4ex}

The expressions for the other three functions and their limiting forms are given below:
\begin{align}
\mbox{Davydov $(q=1, A=a)$:}\hspace{7.5ex} \nonumber\\
\underline{\lambda}_N &=\hspace{3ex} X\left [1-A\cos\theta_N\right ]^{-1}   \hspace{8ex}\\
\noalign{\vspace{0.6ex}}
&\xrightarrow[Z\to 0]{}X\left [1-\,{\frac {A{\pi }^{2}}{2(1-A)}}{Z}^{2}+ {\cal O}(Z^{3}) \right ]\\
\noalign{\vspace{1.4ex}}
\mbox{W. Kuhn $(q=\textstyle\frac{1}{2}, B=b)$:}\hspace{8ex} \nonumber\\
\underline{\lambda}_N &=\hspace{3ex} X\left [1-B\cos\theta_N\right ]^{-1/2}\hspace{8ex}\\
\noalign{\vspace{0.6ex}}
&\xrightarrow[Z\to 0]{}X\left [1-\,{\frac {B\pi^{2}}{4(1-B)}}{Z}^{2} + {\cal O}(Z^{3})\right ]\\
\mbox{Huzinaga and Hasino (H\&H):}\hspace{2ex} \nonumber\\
\underline{\lambda}_N &=  2X \left [1+\sqrt{1+P\sin^2(\pi/N_\pi)}   \right ]^{-1}\\
\noalign{\vspace{0.6ex}}
&\xrightarrow[Z\to 0]{}X\left [1-\textstyle\frac{1}{8}\,P{\pi }^{2}{Z}^{2}+ {\cal O} \left( {Z}^{4} \right)  \right ]
\end{align}

\newpage
\mbox{}

\noindent
{\bf Appendix D: $\boldsymbol{\alpha}$-oligothiophenes}\\\label{page:AppDstart}

Seixas de Melo and coauthors (SdM\&Co), who extended HK's model, applied eq~\ref{eq:HK3p} to spectral data for a series of seven $\alpha$-oligothiophenes ($\alpha N, N=1,2,\ldots 7$); they adjusted all three parameters ($Y$, $U$ and $l$), plotted the fits obtained for the energies of the $S_1$ and $T_1$ states, and noted with satisfaction that the ``adjustments made gave perfect fits" \cite{SdM1999}.

\begin{table}[h!]
  \centering
  \caption{Results obtained by applying ten fitting formulas (with three or two parameters) to spectral data for solutions of $\alpha$-polythiophenes ($N=1\mbox{--}7$) in dioxane}
  \label{tab:SdM02}
  \begin{tabular}{l|c||r}
  \end{tabular}
\vspace*{-6ex}
\end{table}
\begin{figure}[h!] 
  \centering
\includegraphics[width=\linewidth]{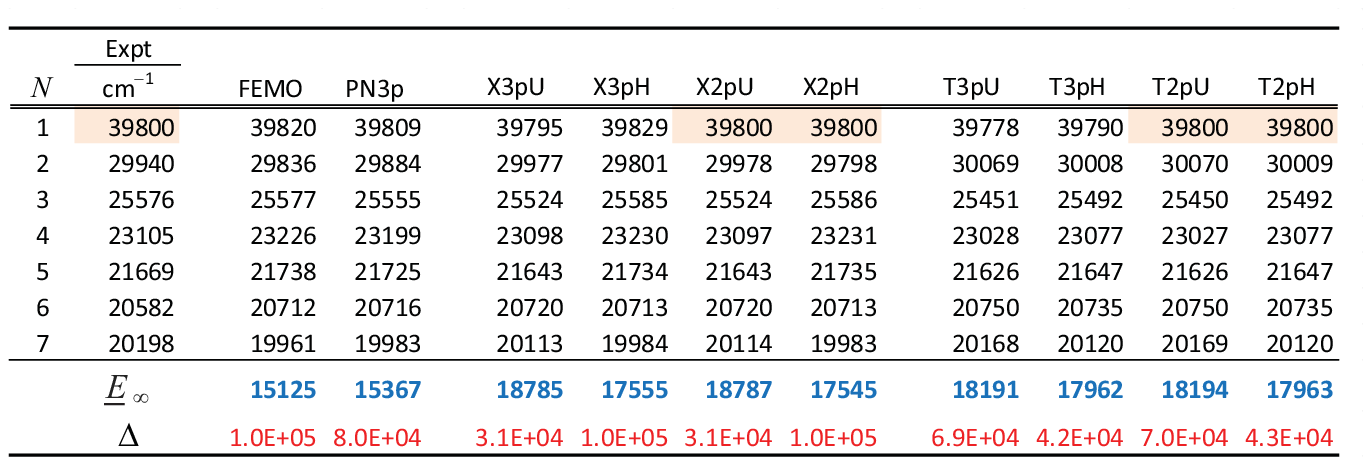}
  \label{fig:SdM02}
\vspace*{-2ex}
\end{figure}
\vspace*{-2ex}
\noindent
Whoever compares the tabulated values of $E_N$ (column 2) with the FEMO predictions of $\underline{E}_N$ (column 3) in Table~\ref{tab:SdM02}, 
\begin{table}[h!]
  \centering
  \caption{Results obtained by applying fitting formulas (with two or fewer parameters) to spectral data for solutions of $\alpha$-polythiophenes ($N=1\mbox{--}7$) in dioxane}
  \label{tab:SdM01}
  \begin{tabular}{l|c||r}
  \end{tabular}
\vspace*{-6ex}
\end{table}
\begin{figure}[h!] 
  \centering
\includegraphics[width=0.7\linewidth]{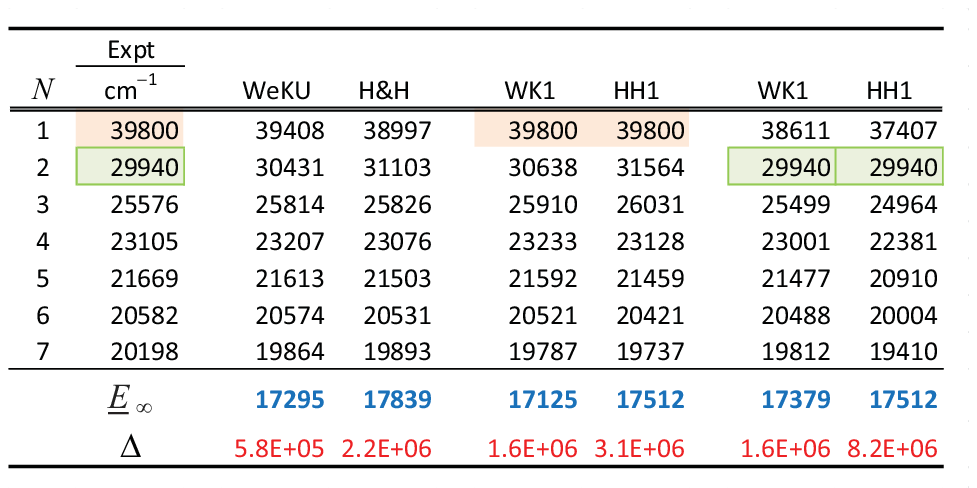}
  \label{fig:SdM01}
\end{figure}
can hardly fail to be impressed by the quality of the fit, but a second-degree polynomial (column 4) and six of the eight other columns show even closer fits; the two functions whose performance is marginally inferior to that of PN3p (but not to that of FEMO) are X3pH (or Hirayma's function) and its two-parameter variant X2pH. The results of a few other fits are also shown in Table~\ref{tab:SdM01}. One entry in each constrained fit has been highlighted to indicate the value of $s$ for which the constraint $\underline{E}_s=E_s$ has been imposed (see Appenix B).

In addition to applying HaKu to their energy levels ($S_1\leftarrow S_0$ and $T_n\leftarrow T_1$), SdM\&Co made an EeZy plot and noted that a linear fit led to $\underline{E}_\infty = 17290\mbox{ cm}^{-1}$ for the limiting energy of the $S_{10}\leftarrow S_{00}$ transition at 293 K, but the $R^2$ value for this fit is only 0.9918; a much better fit, with  $R^2=0.9997$, is obtained when PN3p (eq~\ref{eq:Poly02}) is used, and the result, $\underline{E}_\infty = (15367\pm 200)\mbox{ cm}^{-1}$ (column 4 in Table~\ref{tab:SdM01}), is rather close to the FEMO prediction, and in serious disagreement with the value predicted by the linear fit (which has a standard error of 440 cm$^{-1}$).

SdM\&Co compared the literature values for polythiophene, $E^{[00]}_{N=?}=2.0\mbox{--2.1 eV}$, with their FEMO result (which corresponds to $\underline{E}_{\infty}^{[00]} =1.88 \mbox{ eV in dioxane}$), and concluded that the discrepancy ``signifies that the effective conjugation length of oligomers of medium size represents the polythiophene". If we suppose the new three-parameter fits to be the most reliable (among the values listed in \mbox{Tables} \ref{tab:SdM02} and \ref{tab:SdM01}), and assign them equal weights, we are led to conclude that $\underline{E}_\infty=18100\pm 300$ cm$^{-1}\approx 2.25 \mbox{ eV}$, which is in fair agreement with the results reported by Gierschner and coauthors \cite{Gierschner2007AdvMat}, who fitted WeKu to their data and found $\underline{E}_{\infty}^{[00]} =2.23 \mbox{ eV (hexane)}$ and 2.19 eV (CH$_2$Cl$_2$).

As regards the contents of Ref.~\cite{SdM1999}, it will not be amiss to point out that application of FEMO and the linear fit to the energy levels of $\alpha N$ is only one, and a comparatively minor, aspect of a scrupulously executed investigation with a wider ambit, and the foregoing comments on the inadequacies of the two fits in no way vitiate the other conclusions drawn by the authors.
\label{page:AppDfinish}

\newpage
\mbox{}

\noindent
{\bf Appendix E: A closer look at FEMO}\\\label{page:AppEstart}

It has been stated in \S~\ref{subsection:FEMO} that eq~\ref{eq:HK3p}, reproduced below
\begin{align}
\underline{E}_N &= Y \left [1-\frac{1}{N_\pi}+ \frac{U}{(N_\pi+l)^2} (N_\pi+1)\right ], \tag{\ref{eq:HK3p}}
\end{align}
reduces, at large values of $N$, to a second-degree polynomial in $Z=1/N$, namely
\begin{align}\label{eq:Approx}
\underline{E}_N &= Y \left [ 1 + \frac{U-1}{p}Z + \frac{U(1-2l)}{4p}Z^2  \right ],
\end{align}
but the reader may not immediately appreciate just how large $N$ must be for the right-hand sides of eq~\ref{eq:HK3p} and eq~\ref{eq:HK3pSer} to be regarded as practically identical. The most effective way to put flesh on these mathematical bones is to take data for a long series, divide it into two or more segments (which may or may not overlap), and seek a fit between eq~\ref{eq:HK3p} and  each segment.

\begin{table}[h!]
  \centering
  \caption{Results obtained by the application of eq~\ref{eq:HK3p} (HaKu) and eq~\ref{eq:Approx} (Approx) to QC results for $E_N$ of oligothiophenes}
  \label{tab:AppE}
  \begin{tabular}{l|c||r}
  \end{tabular}
\vspace*{-4ex}
\end{table}
\begin{figure}[h] 
  \centering
\includegraphics[width=0.675\linewidth]{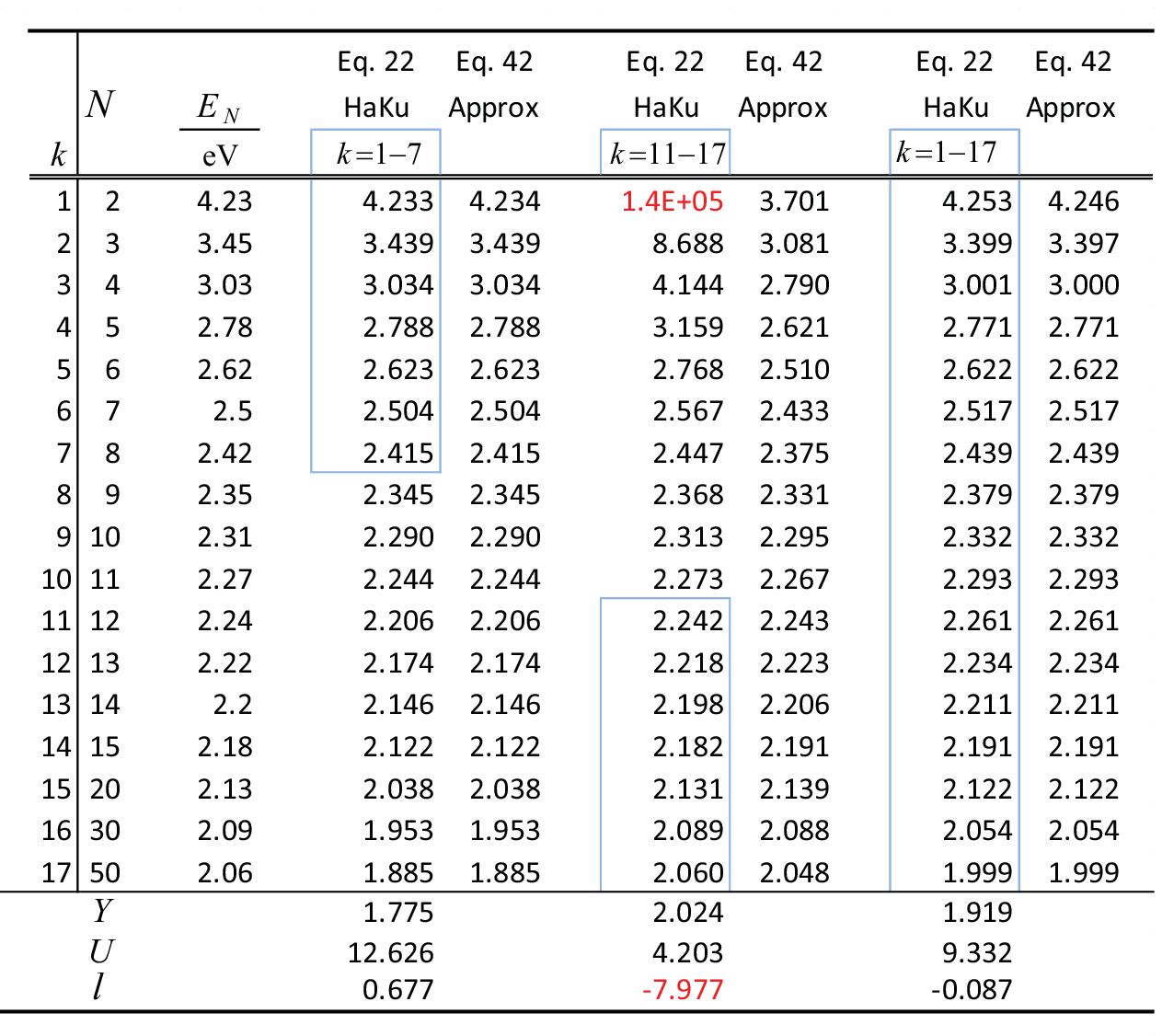}
  \label{fig:AppendixE}
\end{figure}

Let us return to a data set that has already been examined (Tables~\ref{tab:Thio3pB} and \ref{tab:Thio3pA}), namely the QC calculations of a 17-membered series of oligothiophenes ($p=4$ here); the reported energies ($E_N$) are shown in column 3 of Table~\ref{tab:AppE}. Column~4 in this table gives the values of $\underline{E}_N$ (for the entire series) found by applying a HaKu fit that used only the first seven members of the series ($N=k+1=2\mbox{--}8$); the optimized values of the adjustable parameters ($Y$, $U$ and $l$) in column 4 were inserted into eq~\ref{eq:Approx} to obtain the numbers which are displayed in column 5. One does not expect the prediction of HaKu to be close to $E_N$ for values of $N$ significantly larger than 8, and one sees, indeed, that the discrepancy between $E_N$ and $\underline{E}_N$ becomes progressively larger as one moves closer to the end of the series. However, the main purpose of the extension is to permit a comparison between the outputs of eq~\ref{eq:HK3p} and eq~\ref{eq:Approx} for the entire series. The same comments apply to the remaining columns in the table, apart from the fact the fit in column 6 was restricted to the last seven oligomers, and that all oligomers were included in the fit shown in column 8; for the reader's convenience, the energies included in each fit have been enclosed within boxes, and the range has been explicitly stated in the third row under the column labeled HaKu. 

The values of $\underline{E}_N$ in columns 4 and 5 of Table~\ref{tab:AppE} are in near-perfect agreement, and one can expect similar concordance whenever $U$ is appreciably larger than unity and $2l$, so that the factors $U-1$ and $U-2l$ in the coefficients of $Z$ and $Z^2$, respectively, are both positive. In sharp contrast to column 4 is the HaKu fit in the sixth column, where $l$ is negative and almost twice as large in magnitude as $U$. The quality of the fit, far from perfect but passable over the segment for which the parameters were optimized, deteriorates when it is extended to shorter oligomers, and predicts an absurd value for $N=2$, because the denominator of the third term on the right-hand side of eq~\ref{eq:HK3p}, $(N_\pi + l)^2$, becomes very small. In the light of the foregoing comments, it is easy to understand not only why, in the HaKu fit covering all oligomers (column 8), $l$ comes out to be close to zero, but also that, though its inclusion serves to improve a particular fit, the parameter $l$ does not have a constant value for a homologous series.

\begin{table}[h!]
  \centering
  \caption{Optimized values of $Y$, $U$ and $l$ found by fitting eq~\ref{eq:HK3p} to different segments of the data displayed in Table~\ref{tab:AppE}}
  \label{tab:AppETab2}
  \begin{tabular}{l|c||r}
  \end{tabular}
\vspace*{-4ex}
\end{table}
\begin{figure}[h!] 
\hspace*{3ex}
  \includegraphics[bb=0 0 341 60,width=12cm,height=2.11cm,keepaspectratio]{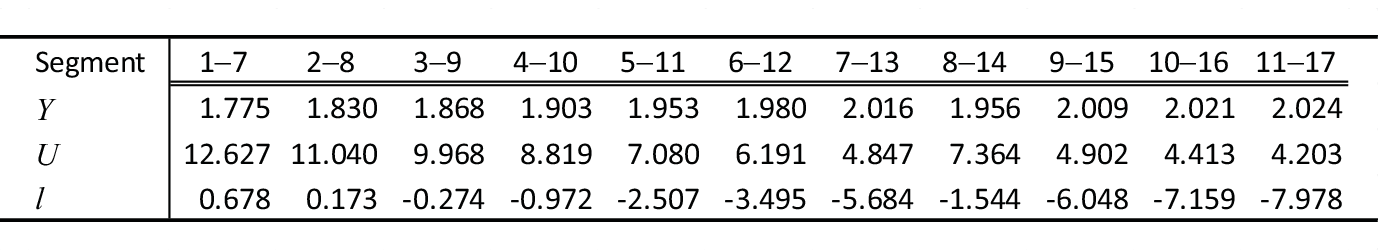}
  \label{fig:AppendixE2}
\end{figure}
The reader can verify that if a HaKu fit is applied successively to a sequence of segments with $k=1\mbox{--}7, 2\mbox{--}8, \ldots$, the data shown in Table~\ref{tab:AppETab2} are obtained. Since the segment with $k=8\mbox{--}14$ does not fit the overall pattern, it will be regarded as a statistical anomaly (an outlier), and ignored; in all the other segments, the three adjustable parameters undergo systematic changes, as the segment moves down the data stream. Since $E_N$ decreases monotonically with $k$, the gradual increase in $Y$ is to be expected and requires no comment, but the larger (and concomitant) decrease in $U$ and $l$ is rather disconcerting, and incompatible with the tenets of HK's model. It seems important to recall at this point, in fairness to HK, that he himself pinpointed the Achilles heel of his model by remarking that the parameter $U$ in eq~\ref{eq:HK1p}, though treated as independent of $N$, decreases with increasing $N$ \cite{Kuhn1949JCP}; the introduction of a third parameter $l$ that takes negative values when $N$ becomes large, serves to compensate for the decline in $U$, but this remedy is insufficient to make FEMO a serious rival to the alternatives that have been presented above.

\label{page:AppEfinish}

The correspondence between FEMO and H\"{u}ckel theory, and the rationale behind placing the ends of the box at one carbon atom beyond either end, have been frequently discussed in the past; I have pointed out elsewhere  \cite{Naqvi1990AnQuim} that there is a tendency
among authors to assume (incorrectly), when they compare a H\"{u}ckel MO with its FEMO counterpart, that both vanish at the end of the box.\\

\newpage

\noindent
{\bf Appendix F:\quad {\em p\/}-Polyphenyls}\\[2ex] \label{page:AppFstart}

\begin{figure}[h] 
  \centering
\includegraphics[width=0.2\linewidth]{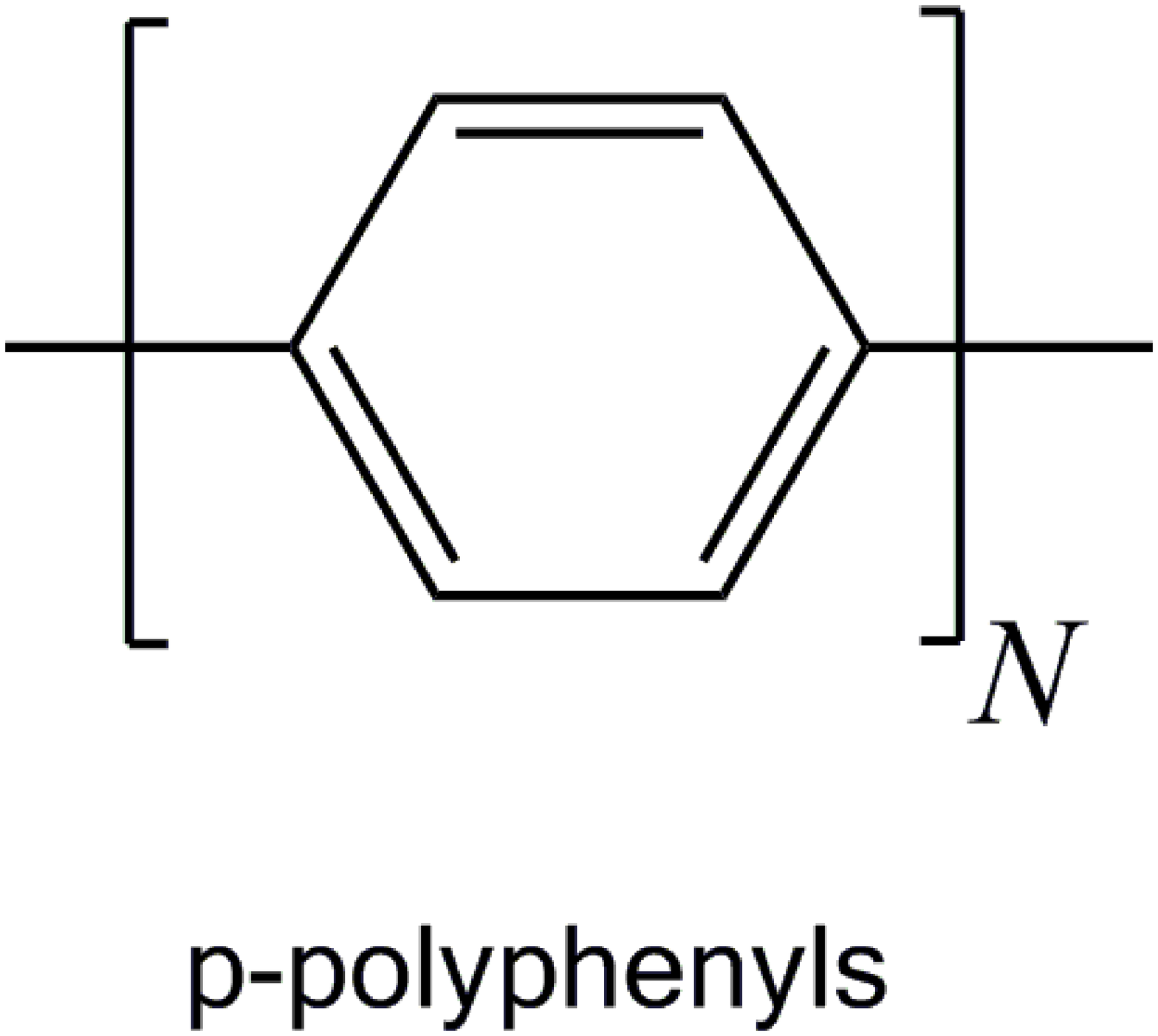}
  \caption{Structure of $p$-polyphenyls}
  \label{fig:Polyphenyles}
\end{figure}

The structure of the $p$-polyphenyls is shown in Figure~\ref{fig:Polyphenyles}. When Davydov \cite{Davydov1948JETP}, Dewar \cite{Dewar1952JCS3} and Hirayama \cite{Hirayama1955JACS} applied their equations to the first five members of the series (with $N=2\mbox{--}6$), all of which have structureless absorption spectra, they 
made use of the $\lambda_N^{[{\rm max}]}$ data reported by Gillam and Hey \cite{GillamHey1939JCS}. Though Platt \cite{Platt1951JCP} had proposed in 1951 that the broad band of biphenyl consists of three overlapping transition (a weak, short-axis polarized transition similar to the ${}^1L_b\leftarrow{}^1A$ in benzene and two stronger transitions, polarized along the long axis, similar to the ${}^1L_a\leftarrow{}^1A$ and ${}^1B_b\leftarrow{}^1A$), further arguments in support of the suggestion were provided much later by Berlman and his collaborators \cite{Berlman1970JCP,Berlman1971JPC}, who also showed that para-substitution by a phenyl ring causes a cross over of energly levels, making the ${}^1L_a\leftarrow{}^1A$ the lowest-energy transition for terphenyl and higher homologues.

\begin{table}[h!]
  \centering
  \caption{Experimental values (in eV) of $E_N^{[00]}$ in the $p$-polyphenyl series and predicted values $\underline{E}_N^{[00]}$ from different fits including (upper part) and excluding (lower part) biphenyl ($N=2$) }
  \label{tab:TablePP}
  \begin{tabular}{l|c||r}
  \end{tabular}
\vspace*{-4ex}
\end{table}
\begin{figure}[h!] 
\includegraphics[width=\linewidth]{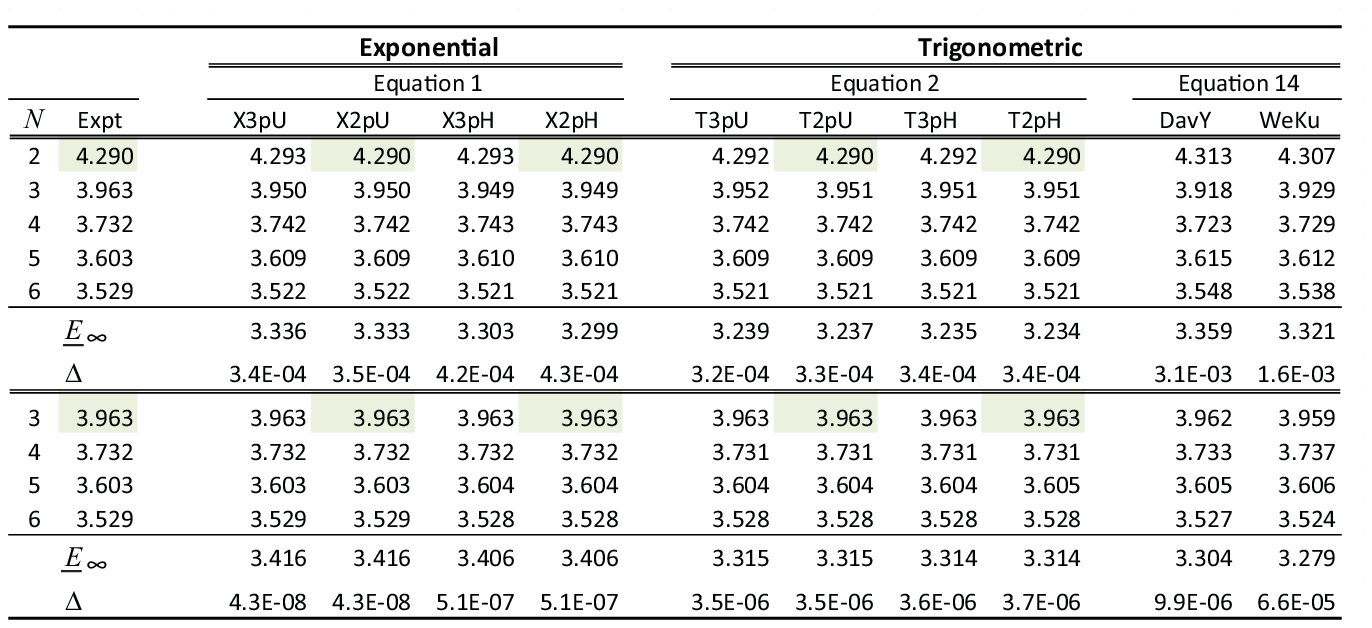}
  \label{fig:AppendixE2}
\vspace*{-12truept}
\end{figure}

Values of $\lambda_N^{[00]}$ (in cm$^{-1}$) reported by Nijegorodov and coworkers \cite{Nijegorodov2000PolyPh} were converted to eV, and used for optimizing the parameters of various fits; the results are displayed in Table~\ref{tab:TablePP}, which is divided horizontally into two parts, the upper one of which includes and the lower excludes biphenyl, the spectroscopically exceptional member of the series; as in other tables in the appendices, one entry in each constrained two-parameter fit has been highlighted to indicate the value of $s$ selected for imposing the constraint.

\begin{figure}[h!] 
  \centering
  \includegraphics[bb=0 0 398 241,width=12cm,height=7.26cm,keepaspectratio]{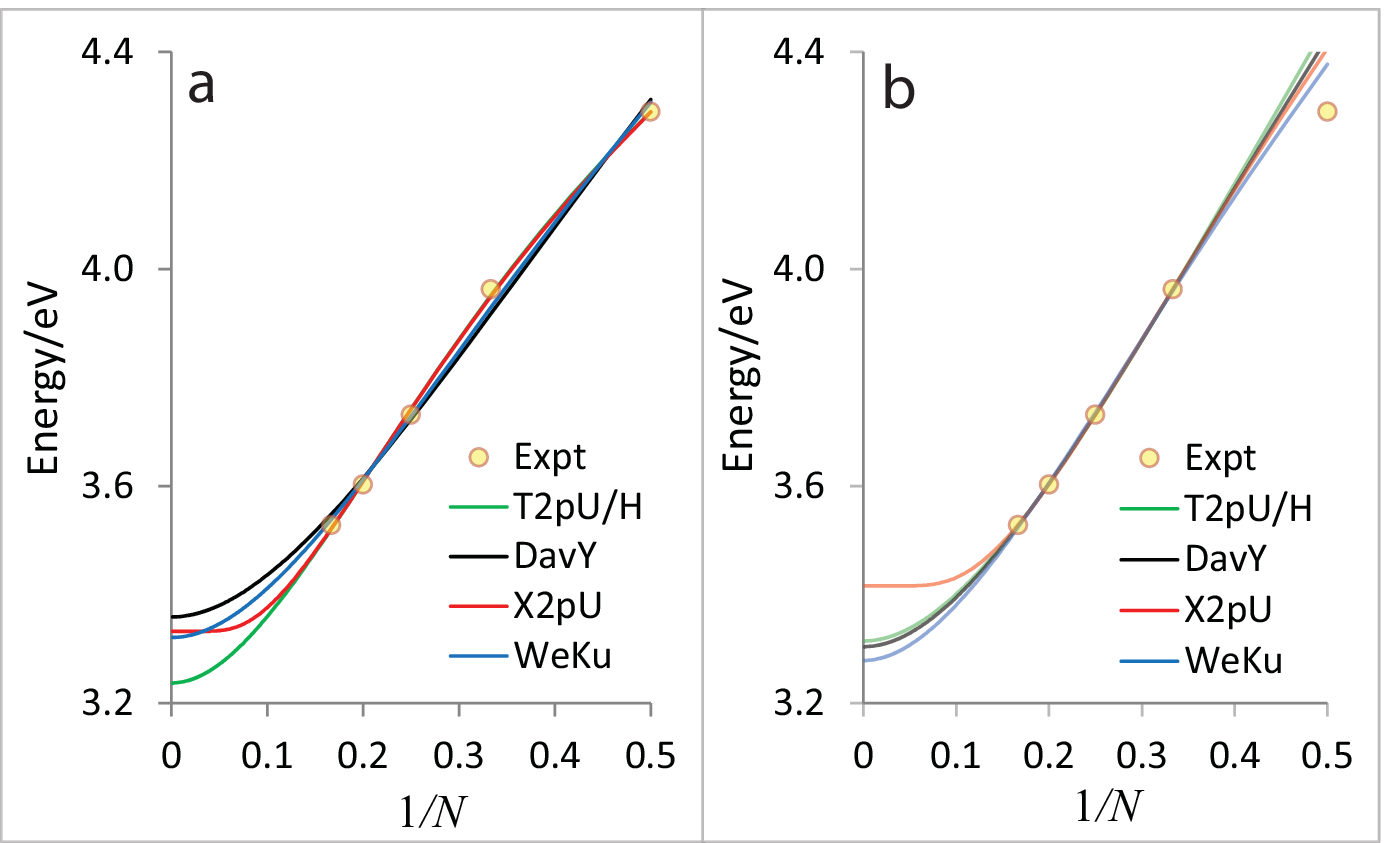}
  \caption{Plots showing experimental data ($E_N^{[00]}$) and values predicted by some two-parameter equations ($\underline{E}_N^{[00]}$); the fits (curves) in panel {\sf a} and {\sf b} are based on data for oligomers with $N=2$--6 and $N=3$--6, respectively.}
  \label{fig:PP-Fits}
\end{figure}

Graphical summaries of some of the results in the upper and lower parts of Table~\ref{tab:TablePP} are shown in Figure~\ref{fig:PP-Fits} (panels {\sf a} and {\sf b}, respectively).

As all the relevant data have been listed in Table~\ref{tab:TablePP}, the table does not need a detailed verbal clarification. However, a few general comments on the differences between the outputs of the four concrete forms of eqs~\ref{eq:MasterX} and \ref{eq:MasterT} (X-equations and T-equations) appear to be in order. Though the X-equations and T-equations predict values of $\underline{E}_N$ that are rather close in the upper half and closer still in the lower, their predictions for $\underline{E}_\infty$ are not so close. One sees also that the exclusion of biphenyl leads to a noticeable improvement in the quality of each fit, but the improvement does not bring the predictions (by X-equations and T-equations) of $\underline{E}_\infty$ into a significantly closer agreement.

One dramatic, though not necessarily significant, consequence of excluding biphenyl from the reckoning is that the results for $\underline{E}_\infty$ fall into two distinct groups, with the X-equations indicating a value close to 3.4 eV, and all the others (T-equations as well as DavY and WeKu) converging around 3.3 eV. 

\label{page:AppFfinish}
\newpage

\noindent
{\bf Appendix G: List of frequently used abbreviations and symbols related to the length of an oligomer}\\[2ex]
\label{page:AppG}
{\small
\hspace*{-4ex}
\renewcommand{\arraystretch}{1.2}
\begin{tabular}{llll}
\toprule
Category	& Abbreviation	& Stands for & Reference  \\
\hline
Symbol 		& $N$      	& number of repeat units        			     \\
		& $N^{=}$      	& number of double bonds along the shortest        	     \\[-1ex]
                & 	     	& path connecting the terminal carbon atoms         		     \\
		& $N_{\pi}=2N^{=}$   	& number of $\pi$ electrons				\\
\hline
Author 		& HK      	& H. Kuhn         	& \cite{Kuhn1949JCP}   		 \\
		& WK      	& W. Kuhn         	& \cite{Kuhn1948Helvetica}   	 \\
		& M\&Co   	& H. Meier and others 	& \cite{Meier1996Liebigs,  Meier1997ActaPol, Meier2002EJOC, Meier2005Ange}\\
		& N\&W		& P. Nayler and M. C. Whiting				&\cite{Nayler1955JCS}	\\
		& SdM\&Co  	& J. Seixas de Melo and others\quad   			& \cite{SdM1999}	\\
\hline
Equation     	& PN2p		& $\underline{E}_N=Y+A_1N^{-1}$				& Equation~\ref{eq:Linear}   \\
		& PN3p		& $\underline{E}_N=Y+A_1N^{-1}+A_2N^{-2}$		& Equation~\ref{eq:Poly02} 	\\
		& HaKu		& HK's equation 			& Equations~\ref{eq:HK1p} and \ref{eq:HK3p}  \\
		& WeKu		& WK's equation		& Equation~\ref{eq:2pTriG} with $q=\textstyle\frac{1}{2}$\\
		& DavY		& Davydov's equation	& Equation~\ref{eq:2pTriG} with $q=1$\\
		& H\&H		& Equation of Huzinaga and Hasino&  Equation~\ref{eq:HandHa}\\
		& X3pU/X3pH	& See Appendix A&\\
		& T3pU/T3pH	& See Appendix A&\\
		& X2pU/X2pH	& See Appendix A&\\
		& T2pU/T2pH	& See Appendix A&\\
\hline
Plot		& EeZy		&  $E_N$-{\em vs.}-$Z$  plot  ($Z\equiv N^{-1}$)     	& Figures~1, 3, 5\\
		& LaZy	&      $\lambda_N$-{\em vs.}-$Z$ plot ($Z\equiv N^{-1}$) 	& Figures~2, 4, 6\\
\bottomrule
\end{tabular}
}

\newpage


{\small
\begin{multicols}{2}


\begin{thebibliography}{99}

\bibitem{LewisCalvin1939CR}
G.~N. Lewis and M.~Calvin.
\newblock The color of organic substances.
\newblock {\em Chem. Rev.}, 25(2):273--328, 1939.

\bibitem{Davydov1948JETP}
A.~S. Davydov.
\newblock Zavisimost chastoty pogloshcheniya sveta para-polifenilami ot chisla
  fenilnykh grupp.
\newblock {\em Zh. Eksp. Teor. Fiz.}, 18(6):515--518, 1948.

\bibitem{Kuhn1948Helvetica}
W.~Kuhn.
\newblock \"{U}ber das absorptionsspektrum der polyene.
\newblock {\em Helv. Chim. Acta}, 31(6), 1948.

\bibitem{Kuhn1949JCP}
H.~Kuhn.
\newblock A quantum-mechanical theory of light absorption of organic dyes and
  similar compounds.
\newblock {\em J. Chem. Phys.}, 17(12):1198--1212, 1949.

\bibitem{Dewar1952JCS3}
M.~J.~S. Dewar.
\newblock {Colour and constitution. Part III. Polyphenyls{,} polyenes{,} and
  phenylpolynes; and the significance of cross-conjugation}.
\newblock {\em J. Chem. Soc.}, (No Vol.):3544--3550, 1952.

\bibitem{Hirayama1955JACS}
K.~Hirayama.
\newblock Absorption spectra and chemical structures. {I--IV}. 
\newblock {\em J. Am. Chem. Soc.}, 77(2):373--379; 379--381; 382--383;
  383--384, 1955.

\bibitem{Huzinaga1957PTP}
S.~Huzinaga and T.~Hasino.
\newblock Electronic energy levels of polyene chains.
\newblock {\em Prog. Theor. Phys.}, 18(6):649--660, 1957.

\bibitem{Tao2006NatureNanotech}
N.~J. Tao.
\newblock Electron transport in molecular junctions.
\newblock {\em Nature Nanotechnology}, 1(3):173--181, 2006.

\bibitem{Zade2010ACR}
S.~S. Zade, N.~Zamoshchik, and M.~Bendikov.
\newblock From short conjugated oligomers to conjugated polymers. Lessons from
  studies on long conjugated oligomers.
\newblock {\em Accounts Chem. Res.}, 44(1):14--24, 2011.

\bibitem{SdM1999}
J.~Seixas~de Melo, L.~M. Silva, L.~G. Arnaut, and R.~S. Becker.
\newblock Singlet and triplet energies of $\alpha$-oligothiophenes: A
  spectroscopic, theoretical, and photoacoustic study: Extrapolation to
  polythiophene.
\newblock {\em J. Chem. Phys.}, 111(12):5427--5433, 1999.

\bibitem{Meier1996Liebigs}
U.~Stalmach, H.~Kolshorn, I.~Brehm, and H.~Meier.
\newblock Monodisperse dialkoxy-substituted oligo(phenyleneethenylene)s.
\newblock {\em Liebigs Annalen}, 1996(9):1449--1456, 1996.

\bibitem{Meier2005Ange}
H.~Meier.
\newblock Conjugated oligomers with terminal donor-acceptor substitution.
\newblock {\em Angew. Chem. Internat. Ed.}, 44(17):2482--2506, 2005.

\bibitem{Gierschner2007AdvMat}
J.~Gierschner, J.~Cornil, and H.-J. Egelhaaf.
\newblock Optical bandgaps of $\pi$-conjugated organic materials at the polymer
  limit: Experiment and theory.
\newblock {\em Adv. Mater.}, 19(2):173--191, 2007.

\bibitem{Torras2012JPCA}
J.~Torras, J.~Casanovas, and C.~Alem\'{a}n.
\newblock Reviewing extrapolation procedures of the electronic properties on
  the $\pi$-conjugated polymer limit.
\newblock {\em J. Phys. Chem. A}, 116(28):7571--7583, 2012.

\bibitem{Meier1997ActaPol}
H.~Meier and H.~Stalmach, U.~Kolshorn.
\newblock Effective conjugation length and UV/vis spectra of oligomers.
\newblock {\em Acta Polymerica}, 48(9):379--384, 1997.

\bibitem{Meier2002EJOC}
H.~Meier and D.~Ickenroth.
\newblock Pentadecamer 2,5-dipropoxy-1,4-phenylenevinylene.
\newblock {\em Eur. J. Org. Chem.}, 2002(11):1745--1749, 2002.

\bibitem{Autschbach2007JCE}
J.~Autschbach.
\newblock Why the particle-in-a-box model works well for cyanine dyes but not
  for conjugated polyenes.
\newblock {\em J. Chem. Educ.}, 84(11):1840, 2007.

\bibitem{Zade2006OrgLett}
S.~S. Zade and M.~Bendikov.
\newblock From oligomers to polymer: Convergence in the Homo-Lumo gaps of
  conjugated oligomers.
\newblock {\em Org. Lett.}, 8(23):5243--5246, 2006.

\bibitem{Nayler1955JCS}
P.~Nayler and M.~C. Whiting.
\newblock {Researches on polyenes. Part III. The synthesis and light absorption
  of dimethylpolyenes}.
\newblock {\em J. Chem. Soc.}, (No Vol.):3037--3047, 1955.

\bibitem{Courant1965IntroCalc}
R.~Courant and F.~John.
\newblock {\em Introduction to Calculus and Analysis}.
\newblock Interscience, New York, 1965.

\bibitem{Glaister1991MathGaz}
P.~Glaister.
\newblock A ``flat" function with some interesting properties and an
  application.
\newblock {\em Math. Gaz.}, 75(474):438--440, 1991.

\bibitem{Slater1939IntroChemPhys}
J.~C. Slater.
\newblock {\em Introduction to Chemical Physics}.
\newblock McGraw-Hill, New York, 1939.

\bibitem{Izumi2003JACS}
T.~Izumi, S.~Kobashi, K.~Takimiya, Y.~Aso, and T.~Otsubo.
\newblock {Synthesis and spectroscopic properties of a series of
  $\beta$-blocked long oligothiophenes up to the 96-mer: Revaluation of
  effective conjugation length}.
\newblock {\em J. Am. Chem. Soc.}, 125(18):5286--5287, 2003.

\bibitem{Naqvi1990AnQuim}
K.~Razi Naqvi.
\newblock On solving the {H}\"{u}ckel equations for linear and cyclic polyenes.
\newblock {\em An. Quim.}, 86(4):337--340, 1990.

\bibitem{GillamHey1939JCS}
A.~E. Gillam and D.~H. Hey.
\newblock {Absorption spectra and structure of compounds containing chains of
  benzene nuclei.}
\newblock {\em J. Chem. Soc.}, (No Vol.):1171--1177, 1939.

\bibitem{Platt1951JCP}
J.~R. Platt.
\newblock Isoconjugate spectra and variconjugate sequences.
\newblock {\em J. Chem. Phys.}, 19(1):101--118, 1951.

\bibitem{Berlman1970JCP}
I.~B. Berlman.
\newblock Identifying the lowest excited singlet state of biphenyl and its
  analogs.
\newblock {\em J. Chem. Phys.}, 52(11):5616--5621, 1970.

\bibitem{Berlman1971JPC}
I.~B. Berlman, H.~O. Wirth, and O.~J. Steingraber.
\newblock Systematics of the electronic spectra of the $p$-oligophenylenes and
  their substituted analogs.
\newblock {\em J. Phys. Chem.}, 75(3):318--325, 1971.

\bibitem{Nijegorodov2000PolyPh}
N. I. Nijegorodov, W. S. Downey, and M. B. Danailov.
\newblock Systematic investigation of absorption, fluorescence and laser
  properties of some $p$- and $m$-oligophenylenes.
\newblock {\em Spectrochim. Acta A}, 56(4):783--795, 2000.

\end{thebibliography}

\end{multicols}
}
\end{document}